\documentclass[12pt,tightenlines,nofootinbib,amsmath,amssymb,aps,prd,balancelastpage]{revtex4-1}

\usepackage{graphicx}
\usepackage[caption=false]{subfig}

\usepackage{amsmath,amssymb}
\usepackage{amsfonts,amssymb,mathrsfs}

\usepackage[bookmarks=false,pdfstartview=FitH]{hyperref}
\usepackage[all]{hypcap}

\usepackage[left=1.1in,right=1.1in]{geometry}

\def\be{\begin{equation}}
\def\ee{\end{equation}}
\def\nn{\nonumber}
\def\f{\frac}

\def\pl{{\rm Pl}}
\def\lp{\ell_\pl}

\def\d{\dot}

\def\bra{\langle}
\def\ket{\rangle}
\def\dd{{\rm d}}

\def\vp{\varphi}

\def\mV{\mathcal{V}}

\newcommand{\SU}{\mathrm{SU}(2)}

\newcommand{\su}{\mathfrak{su}(2)}
\newcommand{\hphi}{\widehat{\varphi}}
\newcommand{\hphid}{{\widehat{\varphi}^\dagger}}
\newcommand{\htphi}{\widehat{\tilde\varphi}}
\newcommand{\htphid}{{\widehat{\tilde\varphi}^\dagger}}

\usepackage{color}

\begin{document}

\pagestyle{plain}

\title{Emergent Friedmann dynamics with a quantum bounce \\ from quantum gravity condensates}

\author{Daniele Oriti} \email{daniele.oriti@aei.mpg.de}
\affiliation{Max Planck Institute for Gravitational Physics (Albert Einstein Institute),\\
Am M\"uhlenberg 1, 14476 Golm, Germany, EU}

\author{Lorenzo Sindoni} \email{lorenzo.sindoni@aei.mpg.de}
\affiliation{Max Planck Institute for Gravitational Physics (Albert Einstein Institute),\\
Am M\"uhlenberg 1, 14476 Golm, Germany, EU}

\author{Edward Wilson-Ewing} \email{wilson-ewing@aei.mpg.de}
\affiliation{Max Planck Institute for Gravitational Physics (Albert Einstein Institute),\\
Am M\"uhlenberg 1, 14476 Golm, Germany, EU}

\begin{abstract}

We study the effective cosmological dynamics, emerging as the hydrodynamics of simple condensate states, of a group field theory model for quantum gravity coupled to a massless scalar field and reduced to its isotropic sector.  The quantum equations of motion for these group field theory condensate states are given in relational terms with respect to the scalar field, from which effective dynamics for spatially flat, homogeneous and isotropic space-times can be extracted.  The result is a generalization of the Friedmann equations, including quantum gravity modifications, in a specific regime of the theory corresponding to a Gross-Pitaevskii approximation where interactions are subdominant.  The classical Friedmann equations of general relativity are recovered in a suitable semi-classical limit for some range of parameters of the microscopic dynamics.  An important result is that the quantum geometries associated with these GFT condensate states are non-singular: a bounce generically occurs in the Planck regime.  For some choices of condensate states, these modified Friedmann equations are very similar to those of loop quantum cosmology.

\end{abstract}

\maketitle

\tableofcontents

\section{Introduction}
\label{s.intro}

One of the main challenges for any fundamental theory of quantum gravity that proposes a candidate for the microscopic degrees of freedom for a quantum space-time is to extract its effective macroscopic physics, which should agree with those of general relativity in some suitable approximation. This is often referred to as the `problem of the continuum' in quantum gravity, since in many approaches the fundamental microscopic degrees of freedom are of a discrete nature (and of no direct spatio-temporal interpretation).  Furthermore, the extraction of effective continuum physics at macroscopic scales is also needed in order to make contact with observations by identifying testable signatures of the proposed microscopic structure of space-time that could be tested at larger scales.  The `problem of the continuum' in quantum gravity is hard, even after one has a promising candidate definition for the microscopic degrees of freedom and their dynamics (which is of course the first hard challenge), because the number of these degrees of freedom in any macroscopic space-time is expected to be extremely large.  This implies that it is necessary to determine the (quantum) dynamics of suitably identified many-body quantum states in the fundamental theory, another difficult task.

One key topic of macroscopic gravitational physics, that on the one hand would benefit from a solid foundation in quantum gravity, and on the other hand is a natural setting where the predictions of quantum gravity effects could realistically be confronted to observations, is cosmology. Therefore the extraction of an effective cosmological dynamics from fundamental quantum gravity models is a particularly interesting and important problem.

In a cosmological context, for homogeneous space-times a few global observables are enough to characterize the macroscopic configurations and dynamics of the space-time and geometry, despite the large number of fundamental degrees of freedom involved. Therefore some kind of truncation, in both the description of the relevant quantum states and of their dynamics, is necessary.  Two main types of truncations are typically considered: coarse-graining (roughly, to use ensembles of microscopic degrees of freedom collectively parametrized by a few variables capturing the relevant macroscopic observables) and symmetry reduction (roughly, to reduce the degrees of freedom of the theory in order to include only the kinematical variables corresponding to these macroscopic observables).  Of course, both strategies involve some approximations to the full theory.

In the coarse-graining approach, the effective dynamics of the collective (cosmologically relevant) observables implicitly capture some effects of the other (coarse-grained away) degrees of freedom; how exactly and how well it manages to do so both depend on the precise coarse-graining scheme as well as on the other approximations that are required to make it work.  From this perspective, cosmology (being the theory of the most global, large-scale degrees of freedom) should arise as a sort of quantum gravity hydrodynamics \cite{CosmoQGhydro}.  In the symmetry reduction approach the relevant cosmological symmetries (e.g., homogeneity and isotropy) are imposed exactly, in order to obtain a system with a minimal number of degrees of freedom, while the others are simply projected out of the resulting theory. If this is done at the microscopic level, it clearly leads to entirely neglecting any effect that the removed degrees of freedom have on the physics of the remaining (cosmological) degrees of freedom. It may thus be seen, in a way, as a particularly brutal form of coarse-graining.  Any form of coarse-graining would then be `better' than such a drastic reduction, as a matter of principle, and especially so in a quantum context.

However, coarse-graining (especially in the full quantum theory) is cumbersome and technically challenging, and in some cases it is certainly true that a symmetry reduction procedure is enough to capture all of the relevant physics in a (sometimes significantly) more straightforward manner.  Whether symmetry reduction does in fact capture the relevant physics can be judged only on a case by case basis; for example, coarse-graining is usually preferred in condensed matter physics, but the easier symmetry reduction works for simple systems like the hydrogen atom.

In the classical gravitational context, symmetry reduction is the main framework used to study the large-scale dynamics of space-time, but even there it is a hotly debated issue whether symmetry reduction gives the same results as coarse-graining gravitational inhomogeneities \cite{Buchert:2011sx, Green:2014aga, Buchert:2015iva}. In a quantum gravitational context, of course, the issue is even murkier and under less control.  Further, even under the assumption that symmetry reduction is enough, it is not clear whether the symmetry reduction should be done at the classical level and one should then quantize the resulting reduced system, or directly at the quantum level.  The two procedures cannot be expected to commute, in general, and much depends on the exact details of the microscopic quantum theory and on its interpretation in relation to continuum general relativity (in particular, whether the full quantum theory is understood as a quantization of general relativity or a quantum theory whose effective description only is given by general relativity).

\bigskip

The above considerations apply to any quantum gravity formalism in which the microscopic degrees of freedom cannot be directly interpreted as quantized continuum geometries, and therefore should be related to continuum geometries through a more careful analysis.

In particular, these considerations are relevant for loop quantum gravity (LQG) \cite{Ashtekar:2004eh, Rovelli:2004tv, Thiemann:2007zz} which, despite its historical roots as a canonical quantization of general relativity in connection variables, ends up with fundamental quantum degrees of freedom which are purely combinatorial and algebraic (e.g., s-knots).  This feature is more apparent in its covariant, or spin foam, formulation, where the same type of combinatorial and algebraic degrees of freedom ---directly obtained from a discretization of the continuum theory--- are used in order to define a prescription for the quantum gravitational path integral.

This is also the case for group field theories (GFTs) \cite{Oriti:2014uga}, which can be seen on the one hand as a second quantized formulation of (the kinematics and dynamics of) loop quantum gravity degrees of freedom \cite{Oriti:2013aqa}, although organized in a different Fock-type Hilbert space, and on the other hand as an enrichment of lattice gravity approaches like tensor models \cite{tensor} (in turn a higher-dimensional generalization of matrix models for 2D gravity, and closely related to the dynamical triangulations formalism for quantum gravity \cite{DT}) by group-theoretic data, i.e., the type of data characterizing quantum geometry in loop quantum gravity.  The GFT framework provides the context for the results presented in this paper.

Due to the nature of the fundamental degrees of freedom in both LQG and GFT, it is reasonable to expect cosmological states to correspond to highly excited states with respect to the natural background independent vacuum state of the theory, which is a `no space'  state (e.g., the Ashtekar--Lewandowski vacuum state \cite{Ashtekar:1993wf} in LQG, and the simple Fock vacuum in GFT).  Thus, to study cosmological space-times in LQG, it will be necessary to only consider a small number of the degrees of freedom of the relevant highly-excited states, and it is clear that some form of coarse-graining is required (perhaps of the simple symmetry reduction type).

In fact, the problem of the continuum (and the related problem of extracting cosmological dynamics from the full theory) is attracting increasing attention both in GFT and in LQG. At the more formal level, it is tied to the study of renormalization flows of quantum gravity models, which is now an active area of research for spin foam models both in their lattice gravity interpretation \cite{biancarenorm} and in their GFT formulation \cite{GFTrenorm,GFTrenorm2}, and to the issue of identifying a suitable phase of the theory in which a continuum geometric rewriting of its dynamics is possible (by defining a non-degenerate geometric vacuum), which has seen considerable progress both in canonical LQG \cite{TimDQG, Koslowski:2011vn, Dittrich:2014wpa, Bahr:2015bra} and in GFT, where it has been suggested that condensate states are natural candidates for non-degenerate geometric vacua.

\bigskip

In the GFT formalism, the hypothesis that the process responsible for the emergence of a continuum geometric space-time is a condensation of the GFT microscopic quanta of space-time (i.e., spin network nodes or equivalently abstract labelled polyhedra) has suggested a promising strategy for extracting an effective cosmological dynamics from the full theory.  Quantum GFT condensate states have been studied in some detail, including both simple condensates corresponding to a gas of uncorrelated GFT quanta (i.e., spin network nodes with no information on their mutual connectivity) and more complex condensates retaining topological information and defined by (infinite) sums over spin network graphs \cite{Gielen:2013kla, Gielen:2013naa} (see also \cite{CondensatesReview} for a detailed overview of the methods and the results).  In all cases, the key feature is that the same `condensate wave function' is assigned to all of the GFT quanta defining the quantum state and hence acts as a collective variable.  This is understood as the coarse-grained quantum gravity analogue of the classical spatial homogeneity condition that characterizes the simplest cosmological space-times.  The effective condensate dynamics can then be extracted directly from the full theory and are given by a non-linear equation for the condensate wave function.  Since the domain of definition of this condensate wave function is isomorphic to the minisuperspace of homogeneous (and potentially anisotropic) geometries, the GFT condensate hydrodynamics has the form of a non-linear extension of loop quantum cosmology and can be studied with similar methods.  Several other results have also been obtained in the study of GFT condensate states, some of which will be reviewed briefly in the following since in this paper we will build on these earlier results.  See \cite{Gielen:2013kla, Gielen:2013naa, Gielen:2014ila, Gielen:2014uga, Sindoni:2014wya} for further details.

\bigskip

Furthermore, the importance of studying many-particle states in quantum gravity for the cosmological sector is also motivated directly from the purely canonical LQG point of view, following from some important insights provided by loop quantum cosmology (LQC) \cite{Ashtekar:2011ni, Banerjee:2011qu, Bojowald:2008zzb}.  In order to clearly describe the input that LQC offers concerning the quanta of geometry in a cosmological space-time, it is helpful to briefly review LQC and explain what is known about the relation between LQC and LQG.

In LQC, symmetry-reduced space-times are quantized in a non-perturbative fashion following the procedures developed in loop quantum gravity.  A lot of progress has been achieved in the field of LQC over the past few years, with the main results being that the big-bang and big-crunch singularities of general relativity are resolved due to quantum gravity effects \cite{Bojowald:2001xe} and are replaced by a bounce \cite{Ashtekar:2006wn}.  To be precise, in LQC the dynamics of a contracting space-time are given by general relativity until the space-time curvature nears the Planck scale, at which point quantum gravity effects become important and cause a bounce which can be seen as a quantum gravity bridge between the contracting and expanding branches of the space-time that can each be treated classically.

Although LQC is based on the same quantization techniques as LQG, LQC is not derived from LQG and the precise relation between LQC and LQG is currently unknown.  This is an important issue to address which is given additional urgency since some predictions of LQC effects have been calculated in a number of cosmological scenarios including inflation \cite{Bojowald:2011iq, Agullo:2013ai, Agullo:2015aba, Bonga:2015xna, Agullo:2015tca} and the matter bounce scenario \cite{WilsonEwing:2012pu, Cai:2014jla} that are of phenomenological interest. It is therefore important to see whether the results of LQC can be provided further support by a proper (if approximate) derivation of LQC from full LQG.  One of the goals of this paper is to shed some light on this problem by studying the dynamics of cosmological states in the GFT reformulation of LQG.

While the issue of the relation of LQC with the full theory has not yet been fully resolved, there has been some important work in this direction including the result that the basis states of LQC can be embedded in the kinematical Hilbert space of LQG \cite{Engle:2013qq} (even though the kinematical Hilbert space of LQC is not the symmetry-reduced kinematical Hilbert space of LQG \cite{Brunnemann:2007du, Hanusch:2013isa}).  However, the main results regarding the relation between LQC and LQG concern their kinematical Hilbert spaces and their basis states, and much less is known about the dynamical sector.

Still, a heuristic argument concerning the relation between LQG and LQC provides key insights which are in close agreement with the discussion above concerning the importance of using highly excited states (with respect to the no space vacuum) in order to study the cosmological sector.  This heuristic relation is based on an observation concerning the kinematical Hilbert space of LQG and the construction of the field strength operator in LQC.

First, quantum states in the kinematical LQG Hilbert space (and in GFT) can be viewed \cite{Barbieri:1997ks, Bianchi:2010gc, Baratin:2010nn, Bianchi:2012wb} as a collection of polyhedra that are glued together across faces with equal areas (although the faces glued together need not have the same shape).  In a typical piece-wise flat interpretation of the resulting discrete structures, space-time curvature is understood to lie on the faces of the polyhedra and the amplitude of the curvature lying on any given face can be evaluated via a holonomy of the Ashtekar-Barbero connection along the edges that form the boundary of that face.

Second, one of the key steps in LQC is the definition of the operator corresponding to the space-time curvature. This is constructed by evaluating the holonomy of the Ashtekar-Barbero connection around a loop of minimal area, as it would be in full LQG.  The choice of the minimal area (rather than some other non-zero%
\footnote{Note that in LQG the limit of Area $\to 0$ is not well-defined since the area has a discrete spectrum and there is no accumulation of eigenvalues near 0.  Thus, since the holonomy around a loop of exactly vanishing area is trivial, in order to appropriately define the field strength operator it is necessary to choose a loop around a finite (although potentially small) area.}
area) is justified by assuming that in the LQG wave function corresponding to the LQC states, there is a large number of polyhedra and that each of their surface areas is given by the minimal area eigenvalue of LQG, $\Delta \lp^2$ (where $\Delta$ is a dimensionless number of order 1) \cite{Ashtekar:2006wn, Ashtekar:2009vc}.

The combination of these two ingredients leads to the following heuristic relation between LQC and LQG. In the simplest, homogeneous and isotropic case, LQC states correspond to wave functions in LQG with a large number of quanta of geometry, or polyhedra, whose surface areas are all given by the minimal non-zero area possible in LQG \cite{Ashtekar:2006wn, Ashtekar:2009vc, Pawlowski:2014nfa}.  This key insight from LQC strongly motivates the use of the family of GFT states we will use in this paper, with the polyhedra specialized for simplicity to tetrahedra.

In other words, the LQG wave function for cosmological space-times, as suggested by LQC, would be a state composed of a large number of identical tetrahedra whose four areas are all equal to $\Delta \lp^2$.  An important point is that these tetrahedra are not only equiareal but also equilateral, a stronger condition.  LQC neglects the connectivity information of the tetrahedra, and without the connectivity information, the LQG wave function of a large number of identical tetrahedra is exactly a simple condensate state.  This is the family of GFT states that we shall consider here.

Now, ignoring the connectivity information of the tetrahedra forming the space-time, for the sake of simplicity but also because this information is not immediately relevant for purely homogeneous and isotropic configurations (and furthermore is ignored in the heuristic relation between LQC and LQG), this type of wave function has only one degree of freedom given by the number of tetrahedra $N$.  This is very restrictive and in order to allow more general states we will remove the requirement that the areas of the faces of each tetrahedron be $\Delta \lp^2$.  Instead, we will assume that the LQG wave functions (in the GFT reformulation) that correspond to cosmological space-times are condensate states composed of a large number of identical equilateral tetrahedra, with no \emph{a priori} restriction on their face areas.  On the basis of these insights, and since condensate states with varying number of quanta are most easily handled in a field theory formalism where there exist creation and annihilation operators, the discussion above suggests that group field theory (GFT) ---which provides a field theory reformulation of LQG, as reviewed in Sec.~\ref{s.gft}--- is the most natural framework to study the cosmological sector of LQG, and that the coarse-grained cosmological dynamics come from the hydrodynamics of GFT condensate states.

\bigskip

Before presenting our results, we point out that there do exist a number of other approaches to the problem of extracting the cosmological sector of LQG, including so-called spin foam cosmology \cite{Bianchi:2010zs, Bianchi:2011ym, Rennert:2013pfa} as well as work on symmetry-reduced canonical loop quantum gravity, either with a large number of spin network nodes \cite{Alesci:2013xd, Alesci:2015nja}  (see also \cite{Lin:2011er} for related earlier work), or just one node \cite{Bodendorfer:2014vea, Bodendorfer:2015hwl}. The approach considered in \cite{Alesci:2013xd, Alesci:2015nja}, in particular, despite the clear difference in strategy and of formalism, considers quantum states quite similar to our GFT condensates. However, all of these alternative approaches assume that the relevant physics is correctly captured by a fixed number of quanta of geometry (which is moreover very small in some cases), an assumption which is seemingly at odds with the lattice-refinement interpretation of LQC which seems to indicate that the number of quanta of geometry increases as the universe expands (or decreases if the space-time is contracting) \cite{Bojowald:2007ra}, as the heuristic relation between LQC and LQG suggests.  See also \cite{Ashtekar:2009vc, Pawlowski:2014nfa, Livine:unp} for further arguments concerning the potential importance for allowing for the number of quanta to change in the cosmological sector of LQG.  In the condensate states studied here there is no restriction on the number of quanta, indeed the number of quanta of geometry will be determined dynamically as it necessarily enters the effective cosmological equations \cite{Gielen:2013kla, Gielen:2013naa,Gielen:2014uga}.  This is one of the important advantages of using GFT condensate states to study the cosmological sector of LQG, in addition to directly addressing the possibility of cosmological space-times being constituted of a large number of quanta of geometry.

\bigskip

In this paper, we will extend previous work on the cosmology of GFT condensates in a number of directions: (i) we work with a GFT that corresponds to (a slight generalization of) the Engle--Pereira--Rovelli--Livine (EPRL) spin foam model of LQG \cite{Engle:2007wy}, which is the most developed Lorentzian model for 4D quantum gravity and is thus a bona fide candidate for the fundamental definition of the quantum dynamics rather than a simple toy model; (ii) we generalize the GFT formalism in order to include a massless scalar field, (iii) in an isotropic restriction, and working with Gross-Pitaevskii condensate states, we extract an effective (relational) cosmological dynamics directly from the fundamental dynamics of the model, showing that it reproduces the Friedmann equations in a classical limit, at least as long as interactions are subdominant; (iv) in the same regime of approximations we derive the leading order quantum gravity corrections to the Friedmann dynamics of the GFT condensate states, which generically predict that the classical big-bang singularity is resolved and is replaced by a non-singular bounce; (v) interestingly, the modified Friedmann equations are qualitatively similar to the effective dynamics of LQC for a simple type of GFT condensate state.  It is worth stressing that, while in this paper we focus on those contributions to the effective dynamics that are expected to be dominant in our approximation (in particular, we consider the interactions as subdominant compared to the kinetic term in the quantum equations of motion as is required in the Gross-Pitaevskii approximation), all of the terms coming from the microscopic dynamics are known and subdominant contributions can be computed explicitly.

More precisely, the outline of the paper is the following.  We start by giving a brief introduction to GFTs in general and reviewing the results of the earlier papers on GFT condensate states \cite{Gielen:2013kla, Gielen:2013naa} in Sec.~\ref{s.gft}, and explain how to include a scalar field in a GFT in Sec.~\ref{s.scalar}.  Then in Sec.~\ref{s.cond} we define the condensate states we take as an ansatz to represent the homogeneous and isotropic sector of LQG, and explain how the scalar field can be used as a relational clock.  In Sec.~\ref{s.dyn} we derive and study the relational Friedmann equations of the GFT condensate states, showing in particular that the big-bang and big-crunch singularities are resolved and are replaced by a bounce, a result analogous to the main qualitative results of LQC.  We end with some comments on the difficulties associated with taking the limiting case of vanishing  matter energy density in Sec.~\ref{s.vac} and a more general discussion in Sec.~\ref{s.disc}.

\section{Basic Elements of the Group Field Theory Formalism}
\label{s.gft}

Group Field Theories (GFTs) \cite{Oriti:2014uga} are a special class of quantum (or statistical) field theories defined over group manifolds (hence the name).  Motivated by quantum Regge calculus \cite{qRC}, dynamical triangulations \cite{DT}, tensor models \cite{tensor}, and loop quantum gravity (LQG) \cite{Ashtekar:2004eh, Rovelli:2004tv, Thiemann:2007zz, Oriti:2014uga, Oriti:2013aqa}, the action for GFTs is typically chosen specifically so that the perturbative expansion of the generating functional generates a sum over diagrams that are dual to simplicial complexes and whose vertices, edges and faces are decorated with group theoretic data.  Then, the Feynman amplitude for each simplicial complex in this sum (once the group theoretic data on the vertices, edges and faces are summed over) corresponds precisely to the quantum gravity path integral evaluated on that particular simplicial complex, assuming appropriate choices are made for the GFT action \cite{Baratin:2010wi}.  The continuum limit is obtained through the sum over all allowed simplicial complexes.

Significantly, GFTs are related to both the canonical and covariant forms of LQG.  On the one hand, GFTs can be understood as a second quantization of canonical LQG \cite{Oriti:2013aqa}, with a different organization of the quantum states so to obtain a Fock space, while the GFT Feynman amplitudes can be equivalently written as spin foam models \cite{Oriti:2014uga}, the covariant form of LQG (see \cite{Perez:2012wv} for an introduction to spin foam models).

Furthermore, as quantum field theories, GFTs admit an operatorial representation in terms of ladder operators acting on conventional Fock spaces.  In this Fock representation, it is clear that GFTs are a second-quantized language for LQG since the GFT quanta are in fact spin network nodes%
\footnote{It is important to stress, however, that GFTs do not refer to background embedding manifolds, nor do they enforce any continuum interpretation of the discrete data entering their quantum states. This is connected to some subtle differences with the conventional LQG constructions of the kinematical Hilbert space, see \cite{Oriti:2013aqa} for a more detailed discussion concerning these points.}.
What will be particularly useful here is that the presence of a Fock space structure renders available some of the powerful methods used in condensed matter physics to construct approximate solutions involving many-body degrees of freedom (which are typically very difficult to access using simple perturbative techniques or attempting to solve directly the many-body Schr\"odinger equation) and to study the corresponding physics \cite{CMT, leggett, Bose}.  These methods may become particularly important in quantum gravity if the continuum limit, in background independent theories of quantum gravity like LQG, does in fact correspond to states with many quanta of geometry as is often expected \cite{Oriti:2014uga, GFTfluid}.

In this section, we will make the above discussion more precise by considering a specific GFT model.  We will first define the field operators in the GFT and explain their geometric interpretation, then introduce the Fock space and operators on the Fock space, and finally present the quantum equations of motion.

\subsection{The Field Operators}
\label{ss.field}

In what follows we will consider a simple bosonic GFT based on the group $\SU^4$ (the most common choice for quantum gravity GFTs). The Fock space is built starting from the single particle Hilbert space $\mathcal{L}^2(\SU^4,\mathbb{C})$ corresponding to four-valent spin network nodes, where the ladder operators are
\begin{equation} \label{field-op}
\hphi(g_{v_1},g_{v_2},g_{v_3},g_{v_4}) , \qquad \hphid(g_{v_1},g_{v_2},g_{v_3},g_{v_4}),
\end{equation}
and these ladder operators are required to be translation invariant under diagonal group multiplication from the right, i.e.,
\be \label{right-inv}
\hphi(g_{v_1}h,g_{v_2}h,g_{v_3}h,g_{v_4}h) = \hphi(g_{v_1},g_{v_2},g_{v_3},g_{v_4}), \qquad \forall \, h \in \SU.
\ee
The identification of this Hilbert space with four-valent spin network nodes is clear: the four $g_{v_i}$ are the parallel transport of an $\su$-valued connection along each of the four links emanating from the same node of a spin network, and the translation invariance encodes the gauge-invariance at the node in spin networks.  For this reason, although the right translation invariance is a global translation in $\SU^4$ it is often referred to as gauge invariance%
\footnote{While it is possible to define this GFT on the coset $\SU^4/\SU_{\mathrm{diag}} \sim \SU^3$, it is convenient to keep the (redundant) $\SU^4$ parametrization as it both simplifies the construction of the theory and makes explicit its interpretation in terms of spin networks (or equivalently in terms of simplicial complexes as shall be explained shortly).}

Note that the choice of four copies of $\SU$ in the definition of the field operators amounts to a restriction to four-valent spin network nodes. This is a minor restriction in the formalism for the sake of simplicity; the theory can be extended, in principle, to include spin network nodes of arbitrary valence \cite{Oriti:2014yla}.

Using the notational shorthand $g_v = (g_{v_1}, g_{v_2}, g_{v_3}, g_{v_4})$ to denote the $\SU^4$ group elements on which the field operator is evaluated, the commutation relations for the ladder operators are
\begin{equation}
[\hphi(g_v) , \hphid(g_w)]  = \int_{\SU} d\gamma \,
\prod_{i=1}^4 \delta(g_{v_i} \gamma g_{w_i}^{-1}),
\end{equation}
which include the correct right gauge invariance.  This shows that the creation operator $\hphid(g_v)$ can be interpreted to create a single four-valent spin network node with data given by $g_v$ up to a gauge transformation on the right.

Equivalently, the same quanta can be interpreted as tetrahedra dual to the four-valent spin network nodes, in which case the four triangular faces of the tetrahedra are dual to the open links leaving the spin network node and are labelled by the same group elements.  To make this dual simplicial interpretation more transparent, it is convenient to rewrite the theory as a field theory on the Lie algebra $\su^4\simeq (\mathbb{R}^3)^4$ via a noncommutative Fourier transform,
\begin{equation}
\htphi(x_v) \equiv \int \dd g_{v} \left(\prod_{i=1}^4 e_{g_{v_i}}(x_{v_i})\right) \hphi(g_v)
\end{equation}
where $e_g(x)$ are noncommutative plane waves and $\dd g_{v} \equiv \prod_{i=1}^4dg_{v_i}$ denotes the Haar measure on $\SU^4$ \cite{Baratin:2010wi, Baratin:2010nn}. The Lie algebra elements should be understood as the normal vectors to each face of the tetrahedron (with their norm corresponding to the area of each face), that in turn are in one-to-one correspondence with the face bivectors.  This rewriting (of both GFT and LQG) depends explicitly on the choice of the quantization map for the flux observables, which will dictate the choice of non-commutative plane waves and their star product \cite{Guedes:2013vi}.

Using the properties of the plane waves $e_g(x)$, it can be verified that the Lie algebra counterpart of the right gauge invariance is the closure constraint for the four faces of the tetrahedron \cite{Baratin:2010nn}.  For this reason, the domain of definition of the field operators $\htphi(x_v), \htphid(x_v)$ is the subspace of $\su^4$ obeying the closure relation
\begin{equation}
x_{v_1} + x_{v_2} + x_{v_3} + x_{v_4} = 0.
\end{equation}
This clearly shows that the quanta created by the field operators can indeed be interpreted as quantum tetrahedra with face bivectors given by the $x_{v_i}$.  Furthermore, it is clear that (at the price of a more complicated theory) it is possible to define a more general GFT than this one which also includes quanta interpreted as higher-valent spin network nodes, or equivalently as convex polyhedra with a greater number of faces.

\subsection{The Fock Space}
\label{ss.fock}

In GFTs, the Fock space is constructed in the standard fashion (assuming Bose statistics) from the `one-particle' Hilbert space described in Sec.~\ref{ss.field} (for details see \cite{Oriti:2013aqa}).  Note that the Fock vacuum represents the state with no spin network nodes, and thus no topological nor geometrical information; the usual spin network states of LQG can be obtained by acting on this vacuum with creation operators convoluted with the desired spin network wave function \cite{Oriti:2013aqa}.

In this second-quantized framework, all operators of interest are generated by polynomials of ladder operators convoluted with suitable kernels where these kernels are the matrix elements of the corresponding first-quantized operators (i.e., the operators of standard canonical LQG like area or volume).  For example, the second-quantized volume operator is (expressed in terms of the Fourier-transformed field operators)
\be
\hat V = \int \dd x_v \,\, \htphid(x_v) \, V(x_v) \, \htphi(x_v),
\ee
where $V(x_v)$ is the volume of a tetrahedron where the normal vectors to each of its faces are $x_v$ and $\dd x_v = \prod_i \dd x_{v_i}$.  This operator gives the total volume of all of the quanta of geometry.

It is also possible to consider more general one-body operators of the type
\begin{equation}
\widehat{O} = \int \dd g_v \dd g_w \,\, \hphid(g_v) \, O(g_v,g_w) \, \hphi(g_w),
\end{equation}
where the field operators do not have the same arguments.  Here $O(g_v,g_w)$ is the LQG matrix element of the operator $O$, evaluated between two nodes decorated with the elements $g_v, g_w$.

There is also a very important new one-body observable that only appears in the second-quantized framework.  This is the number operator
\begin{equation}
\hat{N} = \int \dd g_v \,\, \hphid(g_v) \, \hphi(g_v),
\end{equation}
which counts the number of quanta present in a given state: its eigenvalues distinguish the N-body sectors of the GFT Fock space.  This is also an example of a purely combinatorial observable that characterizes the underlying graph of spin network states, regardless of the attached group theoretic data, which is a type of observable not usually considered in the standard LQG framework.

More general N-body operators that act on more than one spin network node (or tetrahedron), or resulting in more than one, can also be constructed and are of obvious interest.  For example, curvature operators that calculate the parallel transport around a loop containing several spin network nodes will generically be of this type.  Of course, the form of this type of operator is more complicated and, since in this paper we are interested in homogeneous degrees of freedom (to describe homogeneous cosmological space-times), we will restrict our attention to one-body operators like those corresponding to the volume; these will be enough to totally characterize the relevant homogeneous degrees of freedom.

\subsection{The Quantum Equations of Motion}
\label{ss.qeom}

The previous two sections present the key concepts underlining the kinematics of GFTs.  Now, the GFT will be fully specified by giving its dynamics, which are encoded in the GFT action $S[\vp, \bar\vp]$.  The action typically includes involving a kinetic term quadratic in the field operators and some interaction terms which are higher order in the field operators.

Given the action, the path integral for the theory can be (formally) defined, and from the path integral formalism the Schwinger-Dyson equations for correlation functions can be derived, which represent the complete (although formal) specification of the quantum dynamics.

The same quantum dynamics can also be given in operator form, starting from the operator corresponding (up to operator ordering ambiguities) to the classical equations of motion written in terms of field operators, the first equation of motion being
\begin{equation} \label{GFTEOM}
\widehat{\frac{\delta S[\vp,\bar\vp]}{\delta \bar\vp(g_v)}} |\Psi\ket  = 0 \; , 
\end{equation}
and the second being the corresponding equation obtained from the variation of the action with respect to $\vp(g_v)$.

As already mentioned above, the GFT action can be defined to reproduce a spin foam model by choosing the action so that the perturbative expansion of the GFT partition function around the Fock vacuum matches the expansion of the spin foam model.  To be precise, the kinetic term and the interaction term in the GFT action are respectively chosen to reproduce the edge and vertex amplitudes of the spin foam model.

For the most common type of spin foam models which are based on simplicial interactions (i.e., where the spin foam vertices are associated to 4-simplices), the only interaction terms are five-valent and therefore the GFT action will have the general form of
\begin{align} \label{gft-action}
S = \int & \dd g_{v_1} \dd g_{v_2}  \bar\vp(g_{v_1}) \vp(g_{v_2}) \, K_2
+ \f{1}{5} \int \left(\prod_{a=1}^5 \dd g_{v_a} \bar\vp(g_{v_a}) \right) \bar{\mathcal{V}}_5 \nn \\ &
+ \f{1}{5} \int \left(\prod_{a=1}^5 \dd g_{v_a} \vp(g_{v_a}) \right) \mathcal{V}_5,
\end{align}
where the complex-valued functions $K_2 := K_2(g_{v_1},g_{v_2})$ and $\mathcal{V}_5 := \mathcal{V}_5(g_{v_1}, g_{v_2}, g_{v_3}, g_{v_4}, g_{v_5})$ depend on the $g_{v_a}$.  Note that here each $g_{v_a}$ denotes 4 group elements $g_{{(v_a)}_i}$ at each spin network node $v_a$.  The notation is the following: indices $a, b, c, \ldots$ label the node and run from $1, 2, \ldots, N$ with $N$ being the total number of nodes (so in the kinetic term $N=2$ and in the interaction term $N=5$), while indices $i, j$ label the links at a given node and take the values $1, 2, 3, 4$ since only four-valent spin-network nodes are considered in this family of GFT models.

Also, for the sake of simplicity only two interaction terms are included in the above action, one of which contains five creation operators and no annihilation operators and the other the opposite.  Clearly, there are four other interaction terms that could be included (with some relations between them necessary in order to ensure that the action be real). For now we will restrict our attention to these two terms as they shall be sufficient to show how to include any relevant interaction term in any further analysis.

The perturbative expansion of this model gives Feynman diagrams dual to four-dimensional simplicial complexes, and the amplitudes are determined by convolutions of the interaction kernels $\mathcal{V}_5$ associated to the 4-simplices with the inverse kernels of the kinetic term (i.e., the propagator of the theory which connects neighbouring 4-simplices).

Finally, for the action \eqref{gft-action} corresponding to a GFT based on simplicial interactions, the first quantum equation of motion is
\begin{align} 
\left[ \int \dd g_{v_2}  \vp(g_{v_2}) \, K_2
+ \int \left(\prod_{a=2}^5 \dd g_{v_a} \bar\vp(g_{v_a}) \right) \bar{\mathcal{V}}_5 \right]
|\Psi\ket = 0,
\end{align}
with the second quantum equation of motion being the adjoint of this equation.

The specific form of $K_2$ and $\mathcal{V}_5$ will depend on the choices in the construction of the GFT, and in this case we will be interested in the GFT based on the EPRL spin foam model, which is the most studied Lorentzian spin foam model.

One of the main principles used in determining the form of $K_2$ and $\mathcal{V}_5$ is that the imposition of the simplicity constraints ---which transform the 4D topological BF field theory into a geometric theory with local degrees of freedom \cite{DePietri:1999bx,simplicity1,simplicity2,Perez:2012wv}--- can be achieved by modifying the kernels of the 4D $SL(2, \mathbb{C})$ BF Ooguri GFT model with conditions that restrict the four bivectors $x_v$ labeling the triangular faces appearing in the simplicial complexes dual to the Feynman diagrams of the theory to be simple bivectors \cite{Baratin:2011hp, GFT-EPRLj}. This is the GFT counterpart of the standard procedure followed in defining spin foam models \cite{Perez:2012wv}, and following this procedure gives the EPRL GFT model.

However, here we will not go through the details of this lengthy procedure; the interested reader is instead referred to \cite{DePietri:1999bx,GFT-EPRLj}.  Instead, it will be sufficient for our purposes to highlight one of the key properties of the GFT based on the EPRL spin foam model.

To do this, it is convenient to work in the spin representation, where the field operators \eqref{field-op} are rewritten via the Peter-Weyl decomposition
\be
\hphi(g_{v_1},g_{v_2},g_{v_3},g_{v_4}) = \!\! \sum_{j_{v_i},m_{v_i},n_{v_i},\iota} \!\!
\hphi^{j_{v_1}j_{v_2}j_{v_3}j_{v_4} \iota}_{m_{v_1}m_{v_2}m_{v_3}m_{v_4}} \,
\mathcal{I}^{j_{v_1}j_{v_2}j_{v_3}j_{v_4} \iota}_{n_{v_1}n_{v_2}n_{v_3}n_{v_4}}
\prod_{i=1}^4 \frac{1}{d(j_{v_i})} D^{j_{v_a}}_{m_{v_i}n_{v_i}}(g_{v_i}),
\ee
where $d_j = 2 j + 1$.  Note while a generic function of $\SU^4$ would have a more general form, the gauge-invariance of the field operators is translated in the spin representation to the presence of the intertwiners $\mathcal{I}$ labeled by $\iota$.

Then, using the shorthand notation
\be
\hphi^{j_v \iota }_{m_v} \equiv \hphi^{j_{v_1}j_{v_2}j_{v_3}j_{v_4} \iota}_{m_{v_1}m_{v_2}m_{v_3}m_{v_4}},
\ee
the general GFT action for the case of simplicial interactions in the spin representation has the form
\begin{align} \label{gft-action-spin}
S = & \, \sum_{\substack{j_{{v_a}_i} \\ m_{{v_a}_i}, \iota_a}} 
\bar\vp^{\, j_{v_1} \iota_1}_{m_{v_1}}
\vp^{j_{v_2} \iota_2}_{m_2}
\,
(\mathscr{K}_2)^{j_{v_1} j_{v_2} \iota_1 \iota_2}_{ m_{v_1}, m_{v_2}} \nn \\ & \,
+ \f{1}{5} \sum_{\substack{j_{{v_a}_i} \\ m_{{v_a}_i}, \iota_a}} \left[
\left(\prod_{a=1}^5 \bar\vp^{\, j_{v_a} \iota_a}_{ m_{v_a}} \right) \bar{\mathscr{V}}_5
+ \left(\prod_{a=1}^5 \vp^{j_{v_a}\iota_a}_{m_{v_a}} \right) \mathscr{V}_5 \right],
\end{align}
where $\mathscr{V}_5 := \mathscr{V}_5(j_{v_1}, \ldots, j_{v_5}, m_{v_1}, \ldots, m_{v_5}, \iota_1, \ldots, \iota_5)$, and of course each $j_{v_a}$ and $m_{v_a}$ represent the four $j$ and $m$ labels colouring the four links leaving the $v_a$ spin network node.

The key property of the GFT based on the EPRL spin foam model is that (i) the kinetic term contains a Kronecker delta between the $j, m$ and intertwiner labels, and (ii) the interaction term contains a Kronecker delta for the $j$ labels colouring the links that meet in the interaction.  Thus, in this case the kinetic term has the simple form
\be \label{eprl-vac-k}
K = \sum_{j_i, m_i, \iota}\bar\vp^{\, j_{v_1} \iota_1}_{m_{v_1}}
\vp^{j_{v_2} \iota_2}_{m_2}
\,
(\mathscr{K}_2)^{j_{v_1}  \iota_1 }_{ m_{v_1}} 
\delta^{j_{v_1} j_{v_2}} \delta_{m_{v_1} m_{v_2}} \delta^{\iota_1 \iota_2}
\ee
and the interaction term is similarly simplified, with the first interaction term having the form
\begin{align} \label{eprl-vac-v}
V = & \, \sum_{j_i, m_i, \iota_i} \bigg[ 
\vp^{j_1 j_2 j_3 j_4 \iota_1}_{m_1 m_2 m_3 m_4}
\vp^{j_4 j_5 j_6 j_7 \iota_2}_{m_4 m_5 m_6 m_7}
\vp^{j_7 j_3 j_8 j_9 \iota_3}_{m_7 m_3 m_8 m_9}
\vp^{j_9 j_6 j_2 j_{10} \iota_4}_{m_9 m_6 m_2 m_{10}}
\vp^{j_{10} j_8 j_5 j_1 \iota_5}_{m_{10} m_8 m_5 m_1}
\nn \\ & \qquad
\times
\tilde{\mathscr{V}}_5(j_1, \ldots, j_{10}; \iota_1, \ldots, \iota_5) \bigg],
\end{align}
while the second interaction term is simply $V^\dag$.  The input from the EPRL model here is in the combinatoric form of the $j$ and $m$ arguments in the field variables which is due to the presence of Kronecker delta functions in the interaction term which have been taken into account in the above expression (in particular, $\mathscr{V}_5 = \tilde{\mathscr{V}}_5\, \prod_i \delta_{m_i,\tilde{m}_i}$).  The presence of these Kronecker deltas will play an important role in what follows, but the remaining details of the functional form of $\mathscr{K}_2$ and $\mathscr{V}_5$ will not be necessary. For their detailed expression we refer to the literature \cite{Perez:2012wv,GFT-EPRLj}, and we only mention here that they encode a specific relation between the $SU(2)$ representation labels that we use as variables here and $SL(2,\mathbb{C})$ representations, as well as a condition of invariance of the same functions under $SL(2,\mathbb{C})$. This  specific relation between $SU(2)$ and $SL(2,\mathbb{C})$ data is the end result of the EPRL prescription for imposing the constraints reducing topological BF theory to gravity, and which are nothing more than the conditions enforcing geometricity of the simplicial structures on which the model is based.

While it is not necessary to give the exact functional form of $\mathscr{K}_2$ and $\mathscr{V}_5$ here, it is nonetheless important to be aware of a number of ambiguities that arise in the definition of GFT (and spin foam) models.  First, there are the standard factor-ordering ambiguities that generically arise in quantum mechanics (note also that in this setting even operators corresponding to momentum space variables, i.e., the discretized fluxes of the triad field, will not necessarily commute amongst themselves).  Second, the specific choices that are made for $K_2$ and $\mV_5$ are motivated by a discretization of (in this case) the Plebanski--Holst action, and some ambiguities necessarily arise that depend on the specific discretization method.  Third, generically, while the implementation of the simplicity constraints alone (which reduce the topological BF theory to gravity) fixes the type of variables that should appear in the theory so that there is a proper discrete geometric interpretation, and may further suggest the amplitudes associated to vertices, additional requirements (e.g., concerning the composition law of transition amplitudes \cite{Oriti:2000hh, Oriti:2002hv, Bianchi:2010fj}, or the renormalizability of the resulting model) are needed in order to determine the amplitudes associated to edges and faces.  Finally, even once the vertex, edge and face amplitudes are determined, it is possible to redefine these in such a fashion so that the total amplitudes remain unchanged, since these last depend only on convolutions of vertex, edge and face amplitudes in specific combinations.

The existence of construction and quantization ambiguities is of course not surprising as they affect the definition of any quantum theory.  It is hoped that ---and there are general reasons to expect this \cite{Cardy}--- (assuming the continuum and semi-classical limits to be properly defined and shown to exist) most of the large-scale properties of the effective description will be insensitive to many of the details of the construction of the microscopic model, and in particular that they will be insensitive to many of the ambiguities listed above. Of course, the extent of this independence from microphysics will depend on the specific class of phenomena of interest.

We will see that, for the family of GFTs and for the particular class of states that will be considered here, on the one hand there will be a certain degree of universality, but on the other hand a number of these subtle choices in the definition of the GFT model will directly affect the effective cosmological dynamics and thus give different predictions.

\section{Coupling to a Scalar Field}
\label{s.scalar}

So far we have presented the simplest incarnation of GFT models, those which encode only (quantum) geometric degrees of freedom.  However, in order to study most physical applications, and to extract concrete predictions from GFTs, it is necessary to introduce matter fields.   This is especially true in the cosmological context where vacuum homogeneous and isotropic space-times are either Minkowski or (anti-)de Sitter.  Furthermore, when matter fields are present, it is possible to define relational observables (which are diffeomorphism-invariant).  On the other hand, in the vacuum setting it is often very difficult to define Dirac observables.

While the inclusion of matter fields in GFTs (and also in spin foam models) has not been extensively explored (although see \cite{matterGFT1,matterGFT2,matterGFT3, matterSF1,matterSF2,matterSF3}), here we will show how it is possible to include a scalar field in a GFT model.

\subsection{The Field Operators and the Fock Space}
\label{ss.field-phi}

The path that we shall follow here is to simply modify the definition of the field operators so that each GFT quanta carries a value of the scalar field.  To be specific, considering the case of a single real scalar field, the field operators become
\begin{equation}
\hphi(g_v) \rightarrow \hphi(g_v, \phi),
\end{equation}
with the domain of the GFT field operators now $\SU^4 \,\times \, \mathbb{R}$.

Clearly, the Fock space for the GFT with a scalar field can be constructed in exactly the fashion as in the vacuum case, except that in this case the one-particle Hilbert space is $\mathcal{L}^2(\SU^4 \times \mathbb{R},\mathbb{C})$.

An immediate consequence of having a scalar field as an argument in the GFT field is that the presence of the scalar field permits the definition of relational observables, i.e., operators evaluated for a specific value of the scalar field.  Relational observables are particularly useful in cosmology, where matter fields can often act as relational clocks that are well-defined and evolve monotonically globally.  Thus it is often possible to avoid the problem of time in quantum gravity by defining physical evolution with respect to the scalar field `clock'.

For example, besides the `total number' operator
\begin{equation}
\hat{N} = \int d\phi \, \dd g_v \, \hphid(g_v,\phi)\hphi(g_v,\phi),
\end{equation}
it is also possible to define the number operator at a fixed value of $\phi$,
\begin{equation}
\hat{N}(\phi) = \int \dd g_v \, \hphid(g_v,\phi)\hphi(g_v,\phi).
\end{equation}
This is a distributional operator, and it can be regulated by smearing it over a small interval $\Delta\phi$.  Clearly, analogous definitions will hold for other relational operators.

The scalar field is also, of course, a true physical degree of freedom.  There are a number of new observables associated to the scalar field, which are the second-quantized counterpart of the standard observables of a scalar field, namely polynomials in the scalar field and its derivatives.  The two fundamental second-quantized operators are the scalar field operator
\begin{equation}
\widehat{\phi} = \int \dd g_v \, d\phi \, \hphid(g_v,\phi) \phi \hphi(g_v,\phi),
\end{equation}
and the conjugate momentum operator (written in the momentum representation of the scalar field)
\be
\widehat{\pi_{\phi}} = \int \dd g_v \, d\pi_\phi \, \hphid(g_v,\pi_\phi) \,
\pi_\phi \, \hphi(g_v, \pi_\phi).
\ee
The $\widehat{\pi_{\phi}}$ can be rewritten in the $\phi$ representation via a Fourier transform,
\begin{equation}
\widehat{\pi_{\phi}} = \f{\hbar}{2i}  \int \dd g_v \, d\phi \, \left[
\hphid(g_v,\phi) \left( \frac{\partial}{\partial \phi} \hphi(g_v,\phi)\right)
- \left(\frac{\partial}{\partial \phi} \hphid(g_v,\phi) \right) \hphi(g_v,\phi) \right],
\end{equation}
here it is given in a manifestly self-adjoint form.  From this it is easy to define the self-adjoint relational conjugate momentum operator
\begin{equation} \label{pi_phi}
\widehat{\pi_{\phi}}(\phi) = \f{\hbar}{2i}  \int \dd g_v \, \left[
\hphid(g_v,\phi) \left( \frac{\partial}{\partial \phi} \hphi(g_v,\phi)\right)
- \left(\frac{\partial}{\partial \phi} \hphid(g_v,\phi) \right) \hphi(g_v,\phi) \right],
\end{equation}
which will play an important role in extracting the cosmological dynamics from the GFT condensate states.

\subsection{The Quantum Equations of Motion}
\label{ss.qeom-phi}

The dynamics for the GFT models with a scalar field will be determined by the GFT action which is to be chosen following the same procedure as in the vacuum case.  We will once again only consider simplicial interactions, and therefore the action will contain a kinetic term together with interaction terms containing five copies of the field variable.

In this section, we will explain what choices are possible in the GFT action in order to appropriately account for the scalar field.  The construction will be such that, for states where the scalar field vanishes the quantum equations of motion will be precisely those obtained in Sec.~\ref{ss.qeom}.  Of course, the same quantization ambiguities present in the vacuum case will also arise when a scalar field is present.

In order to determine how the scalar field should appear in the kinetic and interaction terms in the action for the case of a GFT with simplicial interactions, it is necessary to understand how a scalar field is discretized on a simplicial discretization of a space-time.

Since a scalar field should be discretized on 0-dimensional elements of the chosen lattice, there are two choices: either real numbers representing the value of the scalar field can be associated to vertices of the simplicial lattice or to nodes of its dual complex.  In the first case, the scalar field will propagate along the edges of the simplicial complex, while in the second case the scalar field will propagate along the links of the dual complex.  While it is reasonable to expect that these two choices lead to equivalent results in the continuum limit, one choice may be better than the other.  First, which choice is most `natural' depends to some extent on the interpretation given to the simplicial complex (and its dual), and second, one choice may give a GFT where calculations are much easier than in the other.

If the basic 4-simplices constituting the simplicial complex are viewed as the finitary substitute of the notion of points in a continuum manifold \cite{finitary} ---in which case the data at each 4-simplex constitute the `local' data in a discrete setting--- then it is most natural to discretize the scalar field on nodes of the dual complex.  In this case a constant value of the scalar field is assigned to the 4D elements of the simplicial lattice, and the gradient terms are given by the difference of the value of the scalar field in neighbouring 4-simplices.  Note that it follows that these discrete gradients are naturally associated to the common boundaries of the neighbouring 4-simplices (i.e., the tetrahedra), and therefore the scalar field propagates along the links of the dual complex. 

Moreover, and perhaps even more importantly, this discretization choice also generically leads to a much simpler extension of the GFT (and spin foam) framework to include scalar matter, as shall become clear.

Locating the discretization on the dual node to the simplicial discretization means that the scalar field has a single value over any 4-simplex.  This directly implies that the five fields in the GFT interaction ---which correspond to the five tetrahedra in a single 4-simplex--- all share the same value of the scalar field.  For this reason the interaction term must impose that the value of the scalar be the same in all of the five interacting fields, in which case the (first) interaction term has the form
\be
V = \int \left( \prod_{a=1}^5 \dd g_{v_a} d \phi_{v_a} \right) \,
\mathcal{V}_5(g_{v_1},\ldots,g_{v_5}; \phi_{v_1})
\, \vp(g_{v_{1}}, \phi_{v_{1}}) \prod_{a=2}^5
\delta(\phi_{v_{a}} - \phi_{v_1})
\vp(g_{v_{a}}, \phi_{v_{a}}),
\ee
where the Dirac $\delta$ distributions are shown explicitly for the sake of clarity.  Clearly, the integrals over $\phi_{v_2}, \phi_{v_3}, \phi_{v_4}$ and $\phi_{v_5}$ are easily evaluated, giving
\be
V = \int \left( \prod_{a=1}^5 \dd g_{v_a} \right) d \phi \,
\mathcal{V}_5(g_{v_1},\ldots,g_{v_5}; \phi)
\prod_{a=1}^5 \vp(g_{v_{a}}, \phi).
\ee
The second interaction term is simply the adjoint of this one.  It is clear that, for GFT models with this interaction term, the scalar field degrees of freedom enter locally in the GFT action (unlike the quantum geometric degrees of freedom that are non-local), and this will simplify a number of calculations.

Since the scalar field propagates along dual links between neighbouring 4-simplices, the gradients of the scalar field will be encoded in the GFT kinetic term, which in general will have the form
\begin{align}
K = \int \dd g_v \dd g_w d\phi_w d\phi_v \,
\bar\vp(g_v,\phi_v) \,
K_2(g_w,g_v; \phi_w,\phi_v)\,
\vp(g_w,\phi_w).
\end{align}

This is the completely general form of a GFT model for quantum gravity coupled to a real scalar field, whose action is given by
\be
S = K + V + V^\dag,
\ee
which follows directly from the discretization strategy described above.

\subsection{A Massless Scalar Field}
\label{ss.massless}

We will now focus on the case of the simplest scalar field coupled to gravity: a minimally coupled, massless scalar field.  While this is a particularly simple case, it is nonetheless enough in order to extract non-trivial cosmological dynamics directly from the GFT model.  Furthermore, a minimally coupled, massless scalar field in a homogeneous space-time acts like a stiff perfect fluid which is of particular interest in early universe cosmology as it is one of the few matter fields that remains relevant (in the sense that its effect on the dynamics of the geometrical degrees of freedom does not become negligible) in the high curvature regime when anisotropies are present.  Therefore, a minimally coupled massless scalar field is of particular interest in studies of quantum gravity effects in cosmology.

An additional advantage of the minimally coupled massless scalar field is that symmetry considerations will be enough to strongly restrict the form of the GFT action.  (Note that for more general cases, whether a non-minimal coupling or the presence of a potential, or both, a more detailed analysis will be necessary in order to determine the appropriate form of the GFT action.)

The matter contribution to the classical (continuum) Lagrangian for a minimally coupled massless scalar field on a curved background is
\begin{equation}
\mathcal{L} = \frac{1}{2} \sqrt{-g} \, g^{\mu\nu}\partial_\mu \phi \partial_{\nu}\phi,
\end{equation}
and this Lagrangian clearly exhibits the shift and sign reversal symmetries
\begin{equation}
\phi(x) \mapsto \phi(x) + c, \qquad \phi(x) \mapsto -\phi(x).
\end{equation}
Note that non-minimal coupling or any non-trivial potential $V(\phi)$ in the scalar field action would kill the first symmetry, and could kill the second, depending the specific form of the non-minimal coupling or potential.

The key step here is that we shall assume that these symmetries are also present at the quantum level in the GFT, and this places strong restrictions on the GFT action.  To be specific, the kinetic term can only depend on the square of scalar field differences $(\phi_1 - \phi_2)^2$, while $\mathcal{V}_5$ must be independent of $\phi$:
\be
K_2(g_w,g_v; \phi_w,\phi_v) = K_2(g_w,g_v; (\phi_w-\phi_v)^2),
\ee
\be
\mathcal{V}_5(g_{v_1},\ldots,g_{v_5}; \phi) = \mathcal{V}_5(g_{v_1},\ldots,g_{v_5}).
\ee
It is clear that any dependance on the field values themselves in the kinetic term or in the interaction kernel would break the shift symmetry, while a dependence of the kinetic term on odd powers of field differences would break the reflection symmetry.

Assuming analyticity of the field variable with respect to the scalar field, it is possible to express the kinetic term as a derivative expansion.  This derivative expansion will be particularly useful in the case where the difference $(\phi_w - \phi_v)$ is small (compared to the Planck mass) in which case the first few terms will provide a good approximation to the full kinetic term.  This derivative expansion can be obtained by rewriting $\phi = \phi_v$ and $u = \phi_w - \phi_v$, and the kinetic term becomes
\be
K = \int \dd g_v \dd g_w \dd\phi \dd u \, \bar\vp(g_v, \phi) K_2(g_v, g_w, u^2) \vp(g_w, \phi + u),
\ee
and the $\vp$ field variable can be Taylor-expanded in its scalar field argument around $\phi$.  Then, after evaluating the integral over $u$, the kinetic term becomes
\be \label{kinetic-exp}
K = \sum_{n=0}^\infty \int \dd g_v \dd g_w \dd\phi \, \bar\vp(g_v, \phi) K_2^{(2n)} (g_v, g_w)
\f{\partial^{2n}}{\partial \phi^{2n}} \vp(g_w, \phi),
\ee
where
\be
K_2^{(2n)} (g_v, g_w) = \int \dd u \, \f{u^{2n}}{(2n)!} \, K_2(g_v, g_w; u^2).
\ee
Note that all of the odd terms in the Taylor expansion do not contribute to the sum in \eqref{kinetic-exp} since the integral over $u$ of an odd power of $u$ multiplying the even function $K_2(g_v, g_w, u^2)$ gives zero.

The functional form of the $K_2^{(2n)} (g_v, g_w)$, which is of course determined by the kinetic term in the GFT action, encodes order by order in derivatives of $\phi$ how the quantum geometric and matter degrees of freedom propagate.  In particular, their exact form could be determined by ensuring that the GFT Feynman amplitudes match term by term the discrete path integral for gravity coupled to a minimally coupled massless scalar field; we leave this analysis for future work.

It will not be necessary to determine the exact functional form of the $K_2^{(2n)} (g_v, g_w)$ for our purposes.  Instead, having ensured that the GFT action has the correct symmetries, we will work in the small derivative limit of the scalar field (i.e., where differences in the scalar field are small compared to the Planck mass), and truncate the kinetic term in the action to the lowest two orders, keeping only the terms corresponding to $n=0$ and $n=1$ in the sum \eqref{kinetic-exp}.  Then, it will be possible to obtain some constraints on their form by studying the dynamics of condensate states of this GFT model ---which as shall be explained shortly, are expected to capture the degrees of freedom of the spatially flat Friedmann-Lema\^itre-Robertson-Walker (FLRW) space-time--- and comparing these dynamics, in the appropriate semi-classical limit, to the Friedmann equations of general relativity.

Finally, here we are interested in the GFT model based on the EPRL spin foam model (whose kinetic and interaction terms in the vacuum case were given respectively in \eqref{eprl-vac-k} and \eqref{eprl-vac-v}) with a minimally coupled massless scalar field.  It is easy to add repeat the procedure outlined in this section starting from the GFT action for the vacuum EPRL model, this gives (in the spin representation)
\be \label{action-exp}
S = K^{(0)} + K^{(2)} + V + V^\dag,
\ee
with
\be
K^{(0)} = \int \dd\phi \sum_{j_i, m_i, \iota} 
\bar\vp^{\, j_v \iota}_{m_v}(\phi)
\vp^{j_v \iota}_{m_v}(\phi) \,
(\mathscr{K}^{(0)}_2)^{j_v \iota}_{m_v},
\ee
\be
K^{(2)} = \int \dd\phi \sum_{j_i, m_i, \iota} 
\bar\vp^{\, j_v \iota}_{m_v}(\phi)
\f{\partial^2}{\partial \phi^2} 
\vp^{j_v \iota}_{m_v}(\phi) \,
(\mathscr{K}^{(2)}_2)^{j_v \iota}_{m_v},
\ee
where the $\vp$ and $\bar\vp$ field variables having the same arguments in the kinetic terms in the action having imposed the Kronecker deltas, and
\begin{align}
V = & \, \sum_{j_i, m_i, \iota_i} \bigg[ 
\vp^{j_1 j_2 j_3 j_4 \iota_1}_{m_1 m_2 m_3 m_4}(\phi)
\vp^{j_4 j_5 j_6 j_7 \iota_2}_{m_4 m_5 m_6 m_7}(\phi)
\vp^{j_7 j_3 j_8 j_9 \iota_3}_{m_7 m_3 m_8 m_9}(\phi)
\vp^{j_9 j_6 j_2 j_{10} \iota_4}_{m_9 m_6 m_2 m_{10}}(\phi)
\vp^{j_{10} j_8 j_5 j_1 \iota_5}_{m_{10} m_8 m_5 m_1}(\phi)
\nn \\ & \qquad
\times
\tilde{\mathscr{V}}_5(j_1, \ldots, j_{10}; \iota_1, \ldots, \iota_5) \bigg],
\end{align}
Note that the kinetic term has been truncated since, as explained above, we are considering the small derivative limit in $\phi$.

A few comments on this GFT model are in order.  First, it was not necessary to assume the presence of any space-time symmetries in order to derive this action; in particular neither homogeneity nor isotropy were imposed.  The simplicity of the GFT action is a result of not only working with a particularly simple matter field (the minimally coupled massless scalar field), but also the chosen discretization where the scalar field is discretized on each 4-simplex.  The simplicity of the GFT classical equations of motion can be understood as the result of coarse-graining the small-scale complexity and obtaining the hydrodynamical equations of motion for the collective behaviour.  This interpretation will be relevant in the following sections.  Second, somewhat surprisingly from a purely formal GFT viewpoint, the minimally coupled massless scalar field enters the GFT action in exactly the same fashion as the standard time coordinate in ordinary quantum field theory.  The presence of the scalar field allows for the definition of a host of new relational observations, but the above observation is stronger: the minimally coupled massless scalar field can be used to define a global relational clock and thus provides a well-defined notion of global time evolution in a diffeomorphism invariant context.  This will be particularly useful in the cosmological context, but it is likely that this will be a powerful tool in a number of other physical settings as well.

\section{GFT Condensates}
\label{s.cond}

As in any interacting quantum field theory, it would be na\"ive to expect to be able to solve the quantum dynamics exactly for realistic GFT models (and note that the situation is potentially worse in the GFT context due to the background independence of GFT models as well as the non-local nature of the quantum geometric interactions).  Instead, the appropriate strategy is to study simplified trial states and look for approximate solutions that may capture the relevant properties of the physical setting of interest.  Then, if these approximate solutions are chosen judiciously, they will provide approximate answers to the physical questions at hand.

In this paper, we will focus on the problem of extracting effective cosmological dynamics from the full quantum gravity formalism, using the GFT model with a minimally coupled massless scalar field based on the EPRL spin foam model described in the previous sections.  In this case, the approximate solutions of interest are GFT condensate states, which are believed to capture the relevant degrees of freedom of the homogeneous FLRW and Bianchi space-times \cite{Gielen:2013kla, Gielen:2013naa}.  We will now summarize the motivation for considering GFT condensate states, and specifically their relevance to the cosmological sector of GFT.  Further details and background information can be found in the literature \cite{Gielen:2013kla, Gielen:2013naa, Gielen:2014ila, Gielen:2014uga, Sindoni:2014wya,CondensatesReview}.

The key idea is that, by definition in a condensate state, all spin network nodes are characterized by the same condensate wave function and in this sense condensate states have a `wave function homogeneity'.  For this reason, condensate states are naturally adapted to the notion of homogeneity, and the coarse-graining procedure which determines the hydrodynamical equations of motion of homogeneous space-times from the microscopic GFT dynamics will be more straightforward in this case than it is in general.  In addition, in GFT condensate states, the condensate wave function for each spin network node stores the intrinsic geometric data of a quantum tetrahedron in 4D, which is isomorphic to the space of homogeneous (and possibly anisotropic) space-times \cite{Gielen:2014ila}.  This will further simplify the coarse-graining procedure.  

In the coarse-graining procedure then, the fact that all of the fundamental GFT quanta are in the same state naturally implements the idea of spatial homogeneity of the continuum (quantum) geometry one can reconstruct from them.  The notion of `continuum' here refers to the fact that an infinite number of fundamental degrees of freedom is captured by the definition of the condensate states; they do not correspond in this sense to any (say, lattice) truncation of the theory but rather to a coarse-graining of it, in which an infinite number of fundamental degrees of freedom is captured by a single collective wave function depending on a small number of collective variables.  This is in full analogy with the use of condensate states in many-body quantum theory, e.g., to extract the relevant macroscopic behaviour of quantum liquids in the condensate phase.  In fact, this particular class of quantum states would be physically (and not only mathematically) appropriate for describing our universe at macroscopic scales in a scenario in which the `geometrogenesis' transition ---between the pre-geometric phase of quantum gravity (which would not admit a continuum description in terms of geometric fields and general relativity) and the geometric one (described by continuous semi-classical geometries)--- is a Bose--Einstein condensation process for the fundamental quantum gravity building blocks of space-time \cite{GFTfluid}.

\newpage

\subsection{The Simplest Condensate States: The Gross--Pitaevskii Approximation}
\label{ss.gp-approx}

Since the aim here is to describe the simplest cosmological space-times ---namely the spatially flat, homogeneous and isotropic Friedmann--Lema\^itre--Robertson--Walker (FLRW) cosmology--- the appropriate degrees of freedom can already be captured in the simplest condensate states.

In particular, since on the one hand the only geometrical observable of interest is the spatial volume%
\footnote{All other geometrical observables in a spatially flat FLRW space-time can be calculated from the spatial volume $V$ and its derivatives with respect to the relational time $\phi$.  For example, the extrinsic curvature is entirely determined by the Hubble rate $H = (3V)^{-1} \cdot dV / dt$ (here $t$ denotes proper time, this can easily be related to the relational time $\phi$ as is explained in Appendix~\ref{app}).}
(as a function of relational time $\phi$) which can easily be calculated without knowing how the spin network nodes are connected amongst themselves, and on the other hand there is no need to encode the spatial curvature in the connectivity of the spin network nodes, it is in fact reasonable to entirely ignore the connectivity of the spin network nodes in the condensate state.

Although this is a drastic approximation, it appears quite reasonable for the spatially flat FLRW space-time.  In addition, this approximation will significantly simplify both the form of the resulting condensate state as well as its quantum equations of motion.

Following the arguments above, we shall neglect the connectivity amongst the spin network nodes, and in this case it is possible to restrict our attention to the relatively simple Gross--Pitaevskii condensate states.  These are coherent states for the GFT field operator,
\begin{equation} \label{def-sigma}
|\sigma\ket = \exp(-\|\sigma\|^2/2) \exp\left( \int \dd g_v\dd\phi \:
\sigma(g_v, \phi) \: \hphid(g_v, \phi) \right) |0\ket,
\end{equation}
where
\begin{equation}
\|\sigma\|^2 = \int \dd g_v \dd\phi \:
|\sigma(g_v, \phi)|^2 = \bra\sigma| \widehat{N} |\sigma\ket.
\end{equation}
As can be seen here, the condensate wave function $\sigma(g_v, \phi)$ is not normalized.  Indeed, its norm captures an important physical quantity, the number of fundamental GFT quanta.  Note that an important property of condensate states is that they are eigenstates of the field operator $\hphi$,
\be \label{cond-eig}
\hphi(g_v, \phi) | \sigma \ket = \sigma(g_v, \phi) | \sigma \ket.
\ee

In order to restrict the condensate wave function $\sigma(g_v, \phi)$ to the space of intrinsic geometries of a tetrahedron in 4D (which as already noted above is isomorphic to the space of spatially homogeneous space-times), an additional restriction is imposed upon the condensate wave function.  Besides the right gauge invariance given in \eqref{right-inv}, the condensate wave function is also taken to be invariant under diagonal left gauge transformations,
\begin{equation} \label{left-inv}
\sigma(h g_{v_1}, h g_{v_2}, h g_{v_3}, h g_{v_4}, \phi) = 
\sigma(g_{v_1},g_{v_2},g_{v_3},g_{v_4}, \phi), \qquad \forall \, h \in \SU.
\end{equation}
%
This choice is rather natural, from a geometric point of view: requiring invariance under diagonal left gauge invariance is equivalent to group-averaging over the relative embedding of the tetrahedron in $\su$, which is not a physically relevant observable.

It is clear that condensate states satisfying the requirement \eqref{left-inv} encode no explicit information about the connectivity of the graphs underlying microscopic states, and therefore also no information about the topology of the space-time itself.  As a result, the appropriate procedure to coarse-grain these states and extract the correct hydrodynamical equations of motion is rather straightforward, but is also less manifest; the coarse-graining is more apparent when working with more general condensate states \cite{Oriti:2015qva}.  While the limitations of the approximation of neglecting the connectivity information should be kept in mind, it is also important to remember that in condensed matter physics the same type of drastic simplification is commonly employed and Gross--Pitaevskii condensate states capture a lot of relevant physics even in their simple form.

Nonetheless, in some settings of physical interest it is to be expected that correlations among GFT quanta may be relevant (in which case quantum fluctuations cannot be neglected), and then the Gross--Pitaevskii approximation will fail.  Indeed, the Gross--Pitaevskii hydrodynamics ---as captured by the simple condensate states of the type \eqref{def-sigma}--- will break down when the interactions between the GFT quanta become sufficiently large.  This can happen either if the coupling constants in front of the interaction terms in the GFT action are big, or when the condensate density (i.e., the modulus of the condensate wave function or equivalently the expectation value of the GFT number operator) grows too large.  We will discuss this point in greater detail in Sec.~\ref{ss.large-density}.

In order to study situations where the interactions become important, it will be necessary to use more complicated condensate states, for example the generalizations defined in \cite{Oriti:2015qva}.  These condensate states contain information about the connectivity of the spin network nodes and hence they are easier to coarse-grain in order to recover the geometry of the resulting space-time; in particular, these generalized condensate states directly provide the topological information of the space-time which is absent in the Gross--Pitaevskii GFT condensate states \eqref{def-sigma}.

Still, even in the Gross--Pitaevskii approximation the condensate wave function $\sigma(g_v, \phi)$ encodes both the number of quanta in the state and the distribution over the possible values of the geometric data of homogeneous (and potentially anisotropic) cosmologies.  Therefore, these condensate states provide a natural point of contact with the usual Wheeler--DeWitt picture of quantum cosmology, and have the additional advantages of (i) not requiring any symmetry reduction, and (ii) including the many-body features of the full Hilbert space of the microscopic theory.  

Furthermore, for these simple condensate states it is straightforward to extract their effective dynamics.
An exact solution to the quantum equations of motion would satisfy all Schwinger--Dyson equations.  However, here we have made an important approximation by assuming that the relevant state is a condensate state of the simplest type.  Therefore, we can only hope for the Schwinger--Dyson equations to be solved approximately.  For this reason we concentrate on only one Schwinger--Dyson equation --- the simplest one which corresponds to the classical equations of motion for the condensate state $\sigma(g_v, \phi)$, which is
\be \label{class-eom}
\bra \sigma | \frac{\delta S[\hphi,\hphid]}{\delta \hphid(g_v)} |\sigma\ket  = 0.
\ee
Note that these equations are non-linear in the condensate wave function and can be non-local on minisuperspace (e.g., if the condensate wave function can be rewritten in terms of minisuperspace variables like the scale factor $a$, the non-linear terms may couple the condensate wave function at different values of $a$).

For the action \eqref{action-exp}, the dynamics for the condensate state \eqref{def-sigma} corresponding to its classical equations of motion is simply (in the spin representation):
\begin{align} \label{gen-cond-eom}
\left[ 
(\mathscr{K}_2^{(0)})^{j_v \iota}_{m_v} + 
(\mathscr{K}_2^{(2)})^{j_v \iota}_{m_v}
\f{\partial^2}{\partial \phi^2} \right] \sigma^{j_v\iota}_{m_v}(\phi)
+ \f{\delta V[\sigma, \bar\sigma]}{\delta 
\bar\sigma^{j_v\iota}_{m_v}(\phi)}
= 0,
\end{align}
with the interaction term contributing a non-linear term of the order of $\bar\sigma^4$, the details of which are unimportant for now.  Also note that there is no sum over $j_v, m_v$ or $\iota$ here.

This equation of motion clearly shows that for the condensate state $|\sigma\ket$, the microscopic dynamics of the GFT quanta can be described hydrodynamically in terms of the collective variable $\sigma(g_v, \phi)$, the condensate wave function.  The hydrodynamic nature of the resulting effective dynamics is the result of the Schwinger--Dyson equations effectively collapsing, due to the simple form of the state $|\sigma\ket$, to the one-particle correlation function (which is exactly $\sigma(g_v, \phi)$ in the condensate approximation \eqref{cond-eig}).  The continuum nature of the equation arises from the fact that $|\sigma\ket$ is given by an infinite sum over numbers of spin network nodes (and hence implicitly over graphs also), and therefore $|\sigma\ket$ is a non-perturbative state with respect to the Fock vacuum.

\subsection{The Large Density Case}
\label{ss.large-density}

As has already been mentioned, in the large density regime ---i.e., when the state has a large number of quanta--- the Gross--Pitaevskii approximation encoded in the simple condensate state \eqref{def-sigma} breaks down.  This is because, while solutions to the complete GFT quantum dynamics satisfy the operator equations \eqref{GFTEOM}, the coherent state ansatz \eqref{def-sigma} only satisfies the expectation value of that same operator equation and thus necessarily fails to capture enough information about the dynamics in regimes where quantum fluctuations and correlations among the GFT quanta become important.

This will be the case, in particular, if the interaction term is large.  To illustrate this point in a bit more detail, let us consider a simple GFT model whose equation of motion is
\begin{equation} \label{large-dens-eom}
\widehat{C}(g_{v_1}) \left| \Phi \right\rangle =
\left[ \hphi(g_{v_1}) + \int \prod_{a=2}^5 dg_{v_a}
V(g_{v_1}, g_{v_2}, g_{v_3}, g_{v_4}, g_{v_5})
\prod_{b=2}^5 \hphid(g_{v_b}) \right]
\left| \Phi \right\rangle = 0 \, .
\end{equation}
While this is a vacuum GFT model, the results obtained here are rather general and also hold for GFT models with a scalar field.

For the coherent state ansatz \eqref{def-sigma}, the expectation value of the above equation of motion gives
\begin{equation}
\sigma(g_{v_1}) + \int \prod_{a=2}^5 dg_{v_a}
V(g_{v_1}, g_{v_2}, g_{v_3}, g_{v_4}, g_{v_5})
\prod_{b=2}^5 \bar\sigma(g_{v_b}) = 0\,,
\end{equation}
and it can be seen that the condensate state does not exactly satisfy the quantum equations of motion \eqref{large-dens-eom}.  This is not surprising since the condensate ansatz is only expected to provide an approximate solution to the equations of motion.  An estimate of the amplitude of the error can be assessed in a number of ways; in the GFT framework, one possibility is to look at the norm of the quantity
\begin{equation}
|\Delta \rangle = \widehat{C}(g_{v_1}) | \sigma \rangle .
\end{equation}
Using the field equations for $\sigma$, the state $|\Delta\ket$ can be rewritten as
\be
|\Delta \ket = \sigma(g_{v_1}) |\sigma\ket + \widehat{V(g_{v_1})} |\sigma\ket
= - V(g_{v_1}) |\sigma\ket + \widehat{V(g_{v_1})} |\sigma\ket,
\ee
with
\be
V(g_{v_1}) = \int \prod_{a=2}^5 dg_{v_a}
V(g_{v_1}, g_{v_2}, g_{v_3}, g_{v_4}, g_{v_5})
\prod_{b=2}^5 \bar\sigma(g_{v_b}),
\ee
\be
\widehat{V(g_{v_1})} = \int \prod_{a=2}^5 dg_{v_a}
V(g_{v_1}, g_{v_2}, g_{v_3}, g_{v_4}, g_{v_5})
\prod_{b=2}^5 \hphid(g_{v_b}).
\ee
Then the norm of $|\Delta\ket$ can easily be calculated, giving
\begin{align}
\langle \Delta | \Delta \rangle & = \langle \sigma |
\left( - \overline{V}(g_{v_1}) + (\widehat{V(g_{v_1})})^{\dagger} \right) 
\left( - {V(g_{v_1})} + (\widehat{V(g_{v_1})}) \right) 
|\sigma \rangle \nonumber \\
& = \langle \sigma |
(\widehat{V(g_{v_1})})^{\dagger} 
(\widehat{V(g_{v_1})})
| \sigma \rangle
-|V(g_{v_1})|^2  
\, .
\end{align}
The second term is exactly the result obtained from normal ordering all of the operators in the expectation values. Therefore, the norm of $|\Delta\ket$ is controlled by the functional of the wave function that is obtained by taking the necessary Wick contractions.  Although the expressions are cumbersome and not particularly illuminating, it is clear that this functional will contain several terms obtained from combining positive powers of the condensate wave function. When $\|\sigma\|^2$ becomes large, then the norm of $|\Delta \rangle$ could become large as well, in general, and at a much faster rate, given the presence of higher powers. In this case the Gross-Pitaevskii approximation would clearly break down.  Whether the Gross-Pitaevskii approximation fails for large $\|\sigma\|^2$ has to be considered on a case by case basis, as it depends on the interplay between the solution and the interaction kernels of the specific GFT model of interest.

This limitation is the analogue of the failure of the dilute gas approximation in the description of the Bose gas whenever the strength of the interactions becomes large with respect to the interatomic distance.  This means that the coherent state approximation can be trusted only in a mesoscopic regime, in which the number of quanta is large enough to admit a continuous geometry approximation but small enough to avoid large deviations from the correct physical state.  In settings where the GFT analogue of the Gross--Pitaevskii ansatz fails, one will have to resort to more involved condensate states (e.g., of the type described in \cite{Oriti:2015qva}) and/or include the dynamics of fluctuations and their coupling to the background condensate dynamics.

\subsection{Restriction to Isotropy}
\label{ss.iso}

In the same spirit of coarse-graining to cosmological observables (rather than symmetry-reduction), we will now impose a further restriction on the form of the condensate wave function so that only isotropic quantities such as the spatial volume or the Hubble rate can be reconstructed from $\sigma(g_v, \phi)$.  Clearly, this restriction on the type of microscopic states allowed will lead to further simplifications in the effective continuum dynamics.  Note that imposing isotropy on the condensate wave function is a restriction on the macroscopic variable that collectively describes all the microscopic configurations entering the definition of the state --- it does not correspond to a symmetry reduction of these microscopic configurations to isotropic ones (this property of the coarse-graining is more apparent for the case of generalized condensate states \cite{Oriti:2015qva}, but also holds in this case.)

Imposing isotropy can be done in a number of different ways.  Since the GFT action for the EPRL spin foam model \eqref{action-exp} is expressed in the spin representation, it is convenient to determine how isotropy is to be imposed in the spin representation.  Since spin network nodes can be interpreted as tetrahedra, it is natural to require that the tetrahedra be as `isotropic' as possible, and therefore we shall define isotropic condensate wave functions to be those that correspond to equilateral tetrahedra (which are the most `isotropic' configurations).  As a result, an isotropic condensate wave function $\sigma$ will depend only on the volume of the tetrahedron or, equivalently, the surface area of one of its faces (as well as on the scalar field $\phi$)%
\footnote{A more precise notion of isotropy should be given in terms of properly defined curvature operators.}.

Since isotropic condensate wave functions are to be those that correspond to equilateral tetrahedra, it is helpful to recall some of their properties.  The classical geometry of an equilateral tetrahedron is entirely described by its four area vectors $\vec{A}_i$, which all have the same norm and are related to each via the discrete subgroup of $O(3)$ called the tetrahedral group.  Equilateral tetrahedra have two properties that will be relevant for the construction of isotropic condensate wave functions: (i) the areas of the faces are equal, $|\vec{A}_i| = |\vec{A}_j| \: \forall \, i,j$ and (ii) for given a total surface area $A_{tot} = \sum_i |\vec{A}_i|$, the volume of the tetrahedron is maximized in the equilateral case.

We can use these two properties to display more clearly what are the true dynamical degrees of freedom that survive in the condensate wave function.  First, using the Peter--Weyl theorem to decompose the wave function into a basis of orthonormal functions given by the Wigner matrices, we see that left and right gauge invariance reduce the wave function to a very simple form:
\begin{equation}
\sigma(g_v;\phi) = \sum_{\{j_v,m_v\}, \iota_l, \iota_r}
\sigma^{j_{v_1} \cdots j_{v_4} \iota_l \iota_r}(\phi) \,
\bar{\mathcal{I}}^{j_1 j_2 j_3 j_4 \iota_l}_{m_1 m_2 m_3 m_4}
\mathcal{I}^{j_1 j_2 j_3 j_4 \iota_r}_{n_1 n_2 n_3 n_4}
\prod_{i=1}^4 \frac{1}{d_{j_{v_i}}} 
D^{j_{v_i}}_{m_{v_i} n_{v_i}}(g_{v_i}),
\end{equation}
in terms of linear combinations of pairs of intertwiners, one associated to the left gauge invariance and one to the right closure condition. The geometric properties are therefore entirely contained in the scalar functions $\sigma^{j_{v_1} \cdots j_{v_4} \iota_l \iota_r}(\phi)$.

The first geometric condition on the isotropic restriction is encoded in the quantum theory by requiring that $\sigma$ must have support only on spin network nodes where the four $j_i$ on each link are equal,
\be
\sigma^{j_{v_1} \cdots j_{v_4} \iota_l \iota_r}(\phi)
 = \sigma^{j \iota_l \iota_r}(\phi) \,
\prod_{i=1}^4 \delta_{j, j_{v_i}}.
\ee
The second condition, the requirement that the volume be maximized in equilateral tetrahedra, is imposed by demanding that for a given $j$ the condensate wave function only has support on the intertwiner that (i) is an eigenvector of the LQG volume operator, and (ii) has the largest eigenvalue.  Denoting this (normalized) intertwiner by $\mathcal{I}^{jjjj \iota_+}_{n_1 n_2 n_3 n_4}$,
\be
\mathcal{I}^{jjjj \iota_+}_{n_1 n_2 n_3 n_4}
= \sum_\iota \alpha_j^\iota \mathcal{I}^{jjjj \iota}_{n_1 n_2 n_3 n_4},
\ee
where the $\alpha_j^\iota$ are determined by the above two conditions together with $\sum_\iota |\alpha_j^\iota|^2 = 1$.

Then, the requirement that the wave function has components only along the equilateral tetrahedra components implies that
\be
\sigma^{j \iota_l \iota_r}(\phi)
= \sigma^{j \iota_l}(\phi) \alpha_j^{\iota_r}.
\ee

Finally, the geometric interpretation of the $\iota_l$ intertwiner (and of the `left flux operators' also) is not immediately clear.  However, some heuristic arguments do provide some guidance.  Since the connectivity in GFTs is imposed (in the spin representation) by requiring that the `left' angular momentum quantum numbers agree between connected links, it may be possible to interpret the left flux operators as the right flux operators parallel-transported from the node of the spin network to the ends of the spin network links.  Since scalar quantities like the volume should behave trivially under parallel transport, this suggests requiring that the volume eigenstate of $\iota_l$ be equal to that of $\iota_r$, and hence that $\mathcal{I}^{jjjj \iota_l}_{n_1 n_2 n_3 n_4} = \mathcal{I}^{jjjj \iota_+}_{n_1 n_2 n_3 n_4}$.  We leave a more detailed study of this condition for future work.

Then, implementing all of the above conditions,
\be
\sigma^{j \iota_l \iota_r}(\phi)
= \sigma_{j}(\phi) \bar \alpha_j^{\iota_l} \alpha_j^{\iota_r}.
\ee

With all these restrictions, coming from geometric reasoning and symmetry arguments, only the form of $\sigma_j(\phi)$ remains to be solved for.  As indicated at the beginning of this section, the condensate wave function (at a given $\phi=\phi_o$) is entirely determined by its dependence on just one geometric quantity, the spin $j$.  For this reason, the only geometric quantities that can be extracted from a condensate wave function of this type are isotropic quantities like the total spatial volume and the Hubble rate, and so \eqref{cond-wf} is a natural candidate to extract the homogeneous, isotropic and spatially flat cosmological sector from quantum gravity.

Finally, for the sake of completeness, note that in the group representation an isotropic condensate wave function has the form
\be
\sigma(g_{v_1},g_{v_2},g_{v_3},g_{v_4}, \phi) = \sum_{j=0}^\infty
\sigma_j(\phi) \, \bar{\mathcal{I}}^{jjjj \iota_+}_{m_1 m_2 m_3 m_4} \,
\mathcal{I}^{jjjj \iota_+}_{n_1 n_2 n_3 n_4} \,
\frac{1}{d(j)^4} \prod_{i=1}^4 D^{j}_{m_i n_i}(g_{v_i}).
\label{cond-wf}
\ee
The form of an isotropic condensate wave function in the flux representation can be obtained via a non-commutative Fourier transform.

\section{Effective Cosmological Dynamics of Isotropic Quantum Gravity Condensates}
\label{s.dyn}

As already outlined in Sec.~\ref{ss.gp-approx}, in the Gross--Pitaevskii approximation, given a condensate wave function for a specific GFT model, it is a straightforward procedure to obtain the (non-linear) equations of motion for that condensate wave function.  In this section, we will derive the equations of motion for the isotropic condensate wave function $\sigma_j(\phi)$ for a GFT model based on the Lorentzian EPRL spin foam model and with a massless scalar field, whose action is given in \eqref{action-exp}.  From these equations of motion it will be possible to extract effective Friedmann equations, expressed in a relational form with respect to the massless scalar field $\phi$.

\subsection{Relational Dynamics}
\label{ss.rel}

As already explained in Sec.~\ref{ss.gp-approx}, the equation of motion for a condensate in the Gross--Pitaevskii approximation is given by \eqref{gen-cond-eom}.  It is worth repeating that the condensate state is expected to be an approximate solution, and therefore not all of the Schwinger-Dyson equations will hold.  Instead, we only impose the first Schwinger-Dyson equation exactly.  For an isotropic condensate wave function of the type \eqref{cond-wf}, and for the GFT action \eqref{action-exp}, based on the EPRL spin foam model and extended in order to include a massless scalar field, this equation has the simple form
\be \label{full-eom}
A_j \partial_\phi^2 \sigma_j(\phi) - B_j \sigma_j(\phi) + w_j \bar\sigma_j(\phi)^4 = 0,
\ee
which are non-linear in $\sigma_j(\phi)$, and the constants $A_j, B_j$ and $w_j$ are obtained from combinations of the kinetic and interaction terms in the GFT action with the non-$\sigma_j(\phi)$ terms in the condensate wave function $\sigma(j_v, m_v, \iota; \phi)$.  To be specific, decomposing the various terms in the action in representations,
\be
A_j =
\sum_{m, \iota_r}
\left(\mathscr{K}_2^{(2)}\right)^{j \iota_r}_m
\bar{\mathcal{I}}^{j \iota_+}_m \mathcal{I}^{j \iota_+}_m
\bar \alpha_j^{\iota_r} \alpha_j^{\iota_r} \, ,
\ee
\be
B_j =
-\sum_{m, \iota_r}
\left(\mathscr{K}_2^{(0)}\right)^{j \iota_r}_m
\bar{\mathcal{I}}^{j \iota_+}_m \mathcal{I}^{j \iota_+}_m
\bar \alpha_j^{\iota_r} \alpha_j^{\iota_r} \, ,
\ee
\be
w_j =
\sum_{\iota_{l_a}, \iota_{r_a}}
\tilde{\mathscr{V}}_5(j, \ldots, j; \iota_{r_1}, \ldots, \iota_{r_5})
\left\{
\begin{array}{ccccc}
\iota_{l_2} &j& \iota_{l_4} & j & \iota_{l_1} \\
j &j& j& j & j \\
j & \iota_{l_3} & j& \iota_{l_5} & j 
\end{array}
\right\}
\prod_{a=1}^5 \alpha_j^{\iota_{l_a}} \bar \alpha_j^{\iota_{r_a}} \, ,
\ee
using a condensed notation in order to avoid writing explicitly all the tensor indices, and here the $15j$ symbol of $\SU$ is that of, e.g., \cite{Yutsis}.

Note that these equations of motion for the condensate wave function can also be obtained from the action
\be \label{sj}
S = \sum_{j=0}^\infty \int \! d\phi \left( A_j |\partial_\phi \sigma_j(\phi)|^2
+ B_j |\sigma_j(\phi)|^2 - \f{1}{5} w_j \sigma_j(\phi)^5 
- \f{1}{5} \bar w_j \bar\sigma_j(\phi)^5 \right),
\ee
obtained by replacing the GFT field in the GFT action by the condensate wave function.  Here it is clear that the scalar field $\phi$ plays the role of a relational time variable.

The condensate equations of motion depend directly on the details of the GFT action, since these determine in part the coefficients $A_j, B_j$ and $w_j$.  It will be possible to constrain their form by requiring that the Friedmann equation be recovered in an appropriate semi-classical limit.

Crucially, the interaction term does not couple $\sigma_j(\phi)$ with different $j$.  This is due to the combination of the isotropic restriction and the form of the EPRL vertex amplitude which contain Kronecker deltas $\delta_{j,j'}$ for all edges that meet in the four-simplex.  Thus, if five equilateral tetrahedra are combined in a four-simplex, and the vertex amplitude is the EPRL one (or one with an analogous property) then it immediately follows that all of the five equilateral tetrahedra must have the same $j$.  This decoupling does not occur generically, even in the isotropic restriction, for other spin foam models, e.g., those like the Baratin--Oriti model \cite{Baratin:2011hp} involving more elaborate fusion coefficients.  For this reason, the interaction term is `local' in the spin label since it has the form $\sim w_j \sigma_j(\phi)^4$ rather than $\sim w_{jklmn} \sigma_k(\phi) \sigma_l(\phi) \sigma_m(\phi) \sigma_n(\phi)$.  Clearly, this significantly simplifies the equations of motion.

As true in general for GFT condensates, we have thus obtained a quantum cosmology-like equation for a cosmological wave function on the space of (isotropic) homogeneous geometries. This equation is however non-linear, as to be expected in a hydrodynamic context, with the non-linearities effectively encoding the microscopic interactions between the fundamental `atoms of space', which are also ultimately responsible for the generation of inhomogeneities at both microscopic and macroscopic scales (see also \cite{Bojowald:2012wi} for a similar construction). 

Before we start analyzing the effective dynamical equations, we point out that, from the symmetries of $S_j$, it is obvious that there is a conserved quantity for every $j$, the `energy' $E_j$ of the condensate wave function $\sigma_j(\phi)$ with respect to the relational time $\phi$,
\be \label{def-ej}
E_j = A_j |\partial_\phi \sigma_j(\phi)|^2
- B_j |\sigma_j(\phi)|^2 + 
\frac{2}{5} \mathrm{Re}\left(
w_j \sigma_j(\phi)^5 \right).
\ee

In addition, in the regime in which the interaction term is small (which is necessary for the Gross-Pitaevskii approximation to hold), the $U(1)$ charge $Q_j$ related to the symmetry $\sigma_j(\phi) \to e^{i\alpha} \sigma_j(\phi)$ emerges as another conserved quantity
\be \label{def-qj}
Q_j = -\f{i}{2} \Big[ \bar\sigma_j(\phi) \partial_\phi \sigma_j(\phi) - \sigma_j(\phi) \partial_\phi \bar\sigma_j(\phi) \Big].
\ee
Note that, following from the definition of the momentum of the massless scalar field, it is easy to check that $\bra\sigma| \hat \pi_\phi(\phi) |\sigma\ket = \hbar \sum_j Q_j$ and therefore $\pi_\phi = \bra\sigma| \hat \pi_\phi(\phi) |\sigma\ket$ is a conserved quantity also in the limit where the Gross-Pitaevskii approximation holds. 

The equation
\be
\f{\partial \pi_\phi}{\partial \phi} = 0
\ee
is exactly the continuity equation in cosmology, for the case of a massless scalar field.  This is a particularly simple example of how the large-scale, coarse-grained effective dynamics can be extracted from the GFT quantum equations of motion for condensate states.  This result is also a first confirmation of the consistency of the identification of the GFT condensate state and an emergent FLRW space-time geometry.

Note that the condition that the interactions be subdominant is required in order to recover the continuity equation for the isotropic condensate state and, as we will show, the Friedmann equations. While this is to some extent only a technical restriction to a regime where simple condensate states can be trusted, it is not unreasonable from a physical point of view. One expects that generic interactions would generate correlations between GFT quanta, and there is no reason to expect these to respect any homogeneity condition, but rather to produce inhomogeneities both the microscopic and macroscopic level. And when inhomogeneities are included in cosmology (even at linear order) the continuity equation is modified.  Note that the heuristic arguments above do not necessarily imply that GFT non-linearities at the level of the hydrodynamic equation encode inhomogeneities (as has been suggested in \cite{steffenInhom}), but this is an interesting hypothesis to explore, especially considering how similar equations (again inspired by BEC theory) have been obtained as an effective description of inhomogeneities in a non-linear extension of (loop) quantum cosmology \cite{Bojowald:2012wi}.   

For the remainder of this paper, we will only consider the limit where the interaction term is much smaller than the linear terms.  This is not because the non-linear case is difficult to solve (in fact, for the simple condensate equations of motion considered here, it is relatively straightforward to study the dynamics of the condensate wave function even in the presence of the non-linear term) but rather because in that limit the Gross-Pitaevskii approximation is expected to fail, in the sense that it cannot be justified from a microscopic point of view since the simple condensate state we use here cannot be expected to be a good approximation to a realistic vacuum of the theory, and it is necessary to consider more complex condensate states than \eqref{def-sigma}.

Therefore, we will study the regime where $|\sigma_j(\phi)|$ is sufficiently small so that the interaction term is subdominant, but at the same time not so small that the hydrodynamic approximation ceases to make sense: after all, $\sum_j |\sigma_j(\phi)|^2$ corresponds to the average number of GFT quanta at the relational time $\phi$, and a large number of quanta is necessary for the hydrodynamic approximation to be valid.  The existence of such a mesoscopic regime depends on the specific form of the GFT action.  To be specific, if the interaction term in the GFT action is sufficiently small compared to the kinetic term, then a mesoscopic regime will exist for some period of relational time.  Indeed, the smaller the interaction term is in relation to the kinetic term, the longer this mesoscopic regime will exist (with respect to the relational clock $\phi$).  The existence of such a mesoscopic regime for appropriate choices of the GFT action has been studied at a phenomenological level in \cite{deCesare:2016rsf}.  In the remainder, we will only consider GFT actions whose interaction term is sufficiently small compared to the kinetic term so that an appropriate mesoscopic regime exists.

In this mesoscopic regime, the equation of motion for $\sigma_j(\phi)$ reduces to
\be \label{linear}
\partial_\phi^2 \sigma_j(\phi) - m_j^2 \sigma_j(\phi) \approx 0,
\ee
with $m_j^2 = B_j / A_j$.

At this point, it is convenient to separate $\sigma_j(\phi)$ into its modulus and phase,
\be
\sigma_j(\phi) = \rho_j(\phi) e^{i \theta_j(\phi)},
\ee
with $\rho_j(\phi)$ and $\theta_j(\phi)$ both assumed to be real, and $\rho_j(\phi)$ to be positive.  From now on, we will drop the argument $\phi$, and denote derivatives with respect to $\phi$ with primes, e.g., $f' := \partial_\phi f(\phi)$.  Then, in terms of $\rho_j$ and $\theta_j$, the equation of motion \eqref{linear} splits into a real and an imaginary part, which are respectively
\be \label{eom-rho1}
\rho_j'' - [m_j^2 + (\theta_j')^2] \rho_j \approx 0,
\ee
and
\be
2 \rho_j' \theta_j' + \rho_j \theta_j'' \approx 0.
\ee
The last equation, coming from the imaginary part of \eqref{linear}, can easily be solved and shows that the combination $\rho_j^2 \theta_j'$ is a constant of the motion, and in fact is precisely the conserved $U(1)$ charge \eqref{def-qj},
\be \label{qj}
Q_j \approx \rho_j^2 \, \theta_j'.
\ee
Note that the other conserved charge, the `GFT energy' for each $j$, also has a simple form,
\be \label{ej}
E_j \approx (\rho_j')^2 + \rho_j^2 (\theta_j')^2 - m_j^2 \rho^2 .
\ee

Finally, using \eqref{qj}, the remaining equation of motion \eqref{eom-rho1} can be rewritten as
\be \label{eom-rho}
\rho_j'' - \f{Q_j^2}{\rho_j^3} - m_j^2 \rho_j \approx 0,
\ee
and this has the form of the equation of motion of a particle in a central potential.  In particular, note that the effective potential diverges as $\rho_j \to 0$; this implies that $\rho_j$ remains non-zero at all times (for non-zero $Q_j$).  This is what will lead to the resolution of the big-bang and big-crunch singularities in the cosmological space-time, as is explained in detail in the next section, so long as the cosmological dynamics are captured by the above equation.

However, before studying the dynamics in more detail and extracting the equations of motion for geometric quantities, it is important to recall the assumptions that were necessary in order to derive \eqref{eom-rho}.  First, we have assumed that a cosmological state in quantum gravity is well-approximated by a simple condensate that in particular ignores connectivity information, which is in general a very important set of dynamical degrees of freedom.  However, in the case of isotropic cosmology we expect these degrees of freedom to play a less important role since the only relevant geometric observables are the spatial volume and its conjugate.  Second, we further imposed that the quanta of geometry in the condensate be isotropic, and we are working in the limit where the scalar field $\phi$ is assumed to evolve slowly.  Finally, we are considering the regime where the interaction term in \eqref{full-eom} is subdominant, and hence where the $\rho_j$ are sufficiently small.

On the other hand, for there to exist a continuum interpretation of the condensate state as a space-time, there must be a large number of quanta of geometry in the condensate state, which requires the $\rho_j$ to be large.  (Also, in order for a consistent continuum geometric interpretation to be valid at least for large total spatial volumes of the universe, a few more conditions are needed, namely that there be a small curvature and a small volume associated to each individual GFT quantum.  These last conditions are not necessary for the mathematical consistency of the condensate approximation, but are necessary to have a clear space-time interpretation for the condensate state.)

A delicate interplay between the values of $\rho_j$ and the coupling constants (and kernels) of the theory is required for the condensate approximation to be valid while at the same time neglecting the interactions.  It is only when all of these assumptions hold that a reliable cosmological interpretation of the condensate state exists and that the effective dynamics extracted here from the full theory can be trusted.

\subsection{Condensate Friedmann Equations}
\label{ss.fried}

The effective dynamics of the GFT condensates is (part of) the hydrodynamics of the GFT model we are studying, and is encoded in an equation for the mean field $\sigma$ (and its complex conjugate) or, in more conventional hydrodynamic form, for a density $\rho$ and a phase $\theta$, which in turn can be decomposed in terms of modes associated to representations $j$.  This type of equation has the form of a non-linear extension of a quantum cosmology dynamics, even though the physical interpretation is different. From this type of equation, just as in (loop) quantum cosmology, it is possible to extract the gravitational dynamics in the form of equations for geometric quantities.  In particular, for homogeneous and isotropic configurations, a natural choice is to derive an effective equation that governs the dynamics of the volume of the universe, coupled to the scalar field.

This can be done in a straightforward fashion in this case starting from the equations of motion for $\rho_j$ obtained in the previous section and relating the spatial volume to the $\rho_j$.  By using the massless scalar field $\phi$ as a relational clock, the resulting equations of motion for $V(\phi)$ can be compared to the Friedmann equations of cosmology, which are presented in the Appendix \ref{app}.

The quantity of interest here is the total volume of the universe in the condensate state, at a given moment of the relational time $\phi$,
\be \label{def-V}
V(\phi) = \sum_j V_j \bar\sigma_j(\phi) \sigma_j(\phi) = \sum_j V_j \rho_j(\phi)^2,
\ee
where $V_j \sim j^{3/2} \lp^3$ is the eigenvalue of the volume operator in canonical loop quantum gravity acting on an equilateral (as defined in Sec.~\ref{ss.iso}) four-valent spin network node in the representation $j$.  (Clearly, it follows from the definition of equilateral spin network nodes that $V_j$ is the largest eigenvalue of the LQG volume operator possible for a node with all $j_i=j$.)  Note that the scaling mentioned here is approximate, and for a detailed analysis it would be necessary to explicitly calculate $V_j$ for each $j$.  However, this will not be necessary here.

A technical comment is also in order here.  The LQG volume operator depends on the Barbero-Immirzi parameter $\gamma$, which only appears in spin foam models after the simplicity constraints have been imposed.  In the GFT models based on spin foam models, the simplicity constraints are imposed in the interaction term in the GFT action, whose effect in the equations of motion has been assumed to be negligible.  However, an operator in GFT can only be interpreted as a geometric operator after simplicity has been imposed.  This is why it is important to remember that we are not ignoring the effect of the interaction term but instead we are considering the case where the contribution of the interaction term to the equations of motion is negligible compared to that of the kinetic terms.  The interaction term is nonetheless present and imposes simplicity, but its contribution to the equations of motion of the condensate wave function is negligible and can be ignored.

Now, given \eqref{def-V}, and using the notation of Sec.~\ref{ss.rel},
\begin{align}
V' \, =\,  2 \sum_j V_j \rho_j' \, \rho_j \,=\, 2 \sum_j V_j \, \rho_j \, {\rm sgn(\rho_j')} \sqrt{ E_j - \f{Q_j^2}{\rho_j^2} + m_j^2 \rho_j^2},
\end{align}
and
\begin{align}
V'' \,=\, 2 \sum_j V_j \Big[ \rho_j'' \, \rho_j + (\rho_j')^2 \Big] \, =\,  2 \sum_j V_j \Big[ E_j + 2 m_j^2 \rho_j^2 \Big].
\end{align}
Both $V'$ and $V''$ depend also on the $w_j$ interaction term in the equations of motion, but the contribution from the interaction term is assumed to be subdominant in the Gross-Pitaevskii approximation and therefore we neglect these terms here.

From the equations above it follows immediately that the generalised Friedmann equations in terms of the relational time $\phi$ are given by
\be \label{gen-cond-friedmann}
\left( \f{V'}{3 V} \right)^2 = \left( \f{2 \sum_j V_j \, \rho_j \, {\rm sgn}(\rho_j') \sqrt{ E_j - \f{Q_j^2}{\rho_j^2} + m_j^2 \rho_j^2}}{3 \sum_j V_j \rho_j^2} \right)^2,
\ee
and
\be
\f{V''}{V} = \f{2 \sum_j V_j \Big[ E_j + 2 m_j^2 \rho_j^2 \Big]}{\sum_j V_j \rho_j^2}.
\ee
These effective Friedmann equations for the GFT condensate include the correct classical limit (i.e., they reproduce the standard Friedmann equations of general relativity, justifying their name), as shall be shown in Sec.~\ref{ss.class}, as well as some quantum corrections coming from the microscopic GFT theory.  Interestingly, some of these corrections have a clear geometric meaning, which shall be discussed shortly.  From these equations, it is possible to solve for the dynamics of the total volume, given some initial state $\sigma_j(\phi_o)$ at an initial time $\phi_o$.

An important point here is that, for the energy density of the massless scalar field, which is defined in terms of the expectation values of scalar field momentum and volume operators as
\be
\rho = \f{\pi_\phi^2}{2 V^2} = \f{\hbar^2 (\sum_j Q_j)^2}{2 (\sum_j V_j \rho_j^2)^2},
\ee
to be non-zero, at least one of the $Q_j$ must be non-zero%
\footnote{The energy density of the massless scalar field $\rho$ ---without an index $j$--- is not to be confused with the amplitude of $\sigma_j(\phi)$ denoted by $\rho_j$, nor with the amplitude $|\sigma|$ of the total condensate wave function $\sigma = \sum_j \sigma_j$.}.
The condition that at least one of the $Q_j$ be non-zero is necessary for the relational dynamics to be well-defined, and also to ensure that the homogeneous and isotropic space-time is an FLRW space-time, not the vacuum Minkowski space-time.

This restriction has important consequences.  Obviously, the condition that at least one of the $Q_j$ be non-zero is a necessary (although not sufficient) condition for the existence of solutions with a good cosmological interpretation, and also for the consistency of the relational description in the first place.  On the other hand, this is not in itself a necessary condition for the mathematical consistency of the condensate dynamics. This means that there may be solutions which do not satisfy this condition, but are still mathematically well-defined and within the regime of validity of the condensate hydrodynamics we are studying.  Therefore, this is an additional requirement beyond the assumptions for a condensate which is necessary for the condensate state to be interpreted as a cosmological space-time.

An open question is whether setting all $Q_j=0$ (but still having large $\rho_j$) gives Minkowski space, in which case the condensate state would correspond to a large space-time although there would be no relational dynamics.  We comment further on the vacuum limit in Sec.~\ref{s.vac}.

Requiring that the energy density of the massless scalar field be non-vanishing has a very important consequence: since at least one $Q_j$ must be non-zero to have a solution that can be interpreted as a cosmological space-time, it follows from \eqref{eom-rho} that at least one $\rho_j$ will always remain greater than zero.  In turn, since $V = \sum_j V_j \rho_j^2$, it follows that $V$ will always remain non-zero.  Therefore, we find that \emph{for all cosmological solutions, the volume will never become zero}.

In this way, the big-bang and big-crunch singularities of classical FLRW space-times that occur generically in general relativity are resolved in the Gross-Pitaevskii GFT condensate states studied here.  The equation of motion for $\rho_j$ \eqref{eom-rho} clearly shows that the individual $\rho_j$ will reach a minimal value at which point they will bounce (and it is clear that there is only a single bounce since $\rho_j$ has only one turning point), and thus the cosmological space-time that emerges from the GFT condensate state is that of a bouncing FLRW space-time.  Note that the bounce occurs when the $\rho_j$ are relatively small, at which time the interactions are weakest, and therefore the near-bounce regime is where the Gross-Pitaevskii approximation used here can be most trusted (for the GFT actions considered here for which an appropriate mesoscopic regime exists); for these GFT actions, the approximation that interactions are small will only fail at large volumes far from the bounce.

In order to see exactly how the singularity is resolved, and better understand the nature of the quantum effects causing this resolution, it is necessary to solve our modified Friedmann equations for $V(\phi)$ for some initial conditions.  Unfortunately, it is difficult to provide an exact solution to these equations of motion for generic initial conditions, but there are two special cases when an explicit solution can be found.

\subsection{Classical Limit}
\label{ss.class}

As already mentioned, the momentum of the scalar field, defined as the expectation value of the operator~\eqref{pi_phi} in the condensate state, is given by $\pi_\phi = \hbar \sum_j Q_j$ and therefore $\pi_\phi$ is a conserved quantity: this is exactly the continuity equation for a massless scalar field in an FLRW space-time.  Therefore, the only other requirement in order to verify that the correct semi-classical limit is obtained is to ensure that the correct Friedmann equation is recovered.

The classical limit of the generalised Friedmann equations is obtained when the Hubble rate is small compared to the inverse Planck time, and this will occur at sufficiently large volumes, i.e., when $\rho_j^2 \gg |E_j| / m_j^2$ and $\rho_j^4 \gg Q_j^2 / m_j^2$ (note that the semi-classical limit is not the limit of large volume, but of small space-time curvature; nonetheless, the space-time curvature decreases as the space-time expands and therefore the dominant term in the Friedmann equation at large volumes is also the dominant term when the space-time curvature is small).  As shall be seen in the next section, the terms containing $E_j$ and $Q_j / \rho_j^2$ can be understood as quantum corrections. 

In this limit, the generalised Friedmann equations become
\be
\left( \f{V'}{3 V} \right)^2 = \left( \f{2 \sum_j V_j \, m_j \, \rho_j^2 \, {\rm sgn}(\rho_j')}{3 \sum_j V_j \rho_j^2} \right)^2,
\ee
and
\be
\f{V''}{V} = \f{4 \sum_j V_j m_j^2 \rho_j^2}{\sum_j V_j \rho_j^2}.
\ee
We immediately see from these equations that, in order to recover the classical Friedmann equations of general relativity in terms of the relational time $\phi$, which are given in Appendix~\ref{app-gr}, (in this specific context where the FLRW space-time emerges as a condensate of isotropic GFT quanta) it is necessary to identify $m_j^2 = 3 \pi G$ for all $j$.  For these values of $m_j$, the GFT condensate dynamics reproduce the classical Friedmann equations of general relativity.  (As an aside, note that while it may be possible, at a specific relational instant $\phi_o$, to choose a different set of values for $m_j$ that also gives the correct limit, this identification will not be preserved by the dynamics and hence the correct classical Friedmann equations would in this case only be recovered in a small neighbourhood of relational time around $\phi_o$.)

The condition that $m_j^2 = 3 \pi G$ is a requirement on the form of the terms $A_j$ and $B_j$ that are determined by the GFT action: if $B_j / A_j \neq 3 \pi G$ for some $j$, then it follows that the correct Friedmann equations are not recovered in the classical limit.  Note also that this should be understood as a definition of $G$ which arises as a hydrodynamic parameter and it is thus a function of the microscopic GFT parameters, and not as an interpretation of the microscopic parameters.  This is an important conceptual point since this identification has no reason to be valid in a generic regime of the dynamics (e.g., for non-condensate GFT states) and may be different in other settings.

So, if all $m_j^2 = 3 \pi G$, then the generalised Friedmann equations of the GFT condensate become, in the classical limit,
\be
\left( \f{V'}{V} \right)^2 = \f{V''}{V} = 12 \pi G,
\ee
which are exactly the Friedmann equations of general relativity for a spatially flat FLRW space-time with a massless scalar field $\phi$, used as a relational time (see Appendix~\ref{app-gr} for details).

The solution to these equations of motion is the standard one of classical general relativity,
\be
V = V_o e^{\pm \sqrt{12 \pi G} \phi},
\ee
as expected, with the sign in the exponent depending on whether the universe is expanding or contracting, and $V_o$ depending on the initial conditions.

\subsection{Single Spin Condensates}
\label{ss.spin}

The other case where the equations of motion for $V(\phi)$ can be solved exactly, and for generic initial conditions, is when only one $\rho_j$ is non-zero, which corresponds to a condensate wave function that is very sharply (infinitely) peaked in $j$,
\be \label{single-spin}
\sigma_j(\phi) = 0, ~{\rm for~all}~ j \neq j_o.
\ee
Then the sum over $j$ in all of the expressions trivializes and an exact solution can be found which includes quantum corrections.

This assumption mirrors the situation that is thought to be relevant in LQC, where there is also the extra assumption that the underlying LQG state consists of a graph with a very large number of nodes and links, and that the spins on all of the links are identical (often chosen to be $j_o = 1/2$).  It follows that in the LQC picture, a cosmological space-time expands or contracts by modifications to the combinatorial structure of the spin network that consist of adding or removing nodes, rather than by changing the spin labels on the spin network; this is analogous to the volume dynamics extracted from the underlying GFT model where changes in $V$ correspond to changes in the number of GFT quanta, rather than transitions between GFT quanta coloured by different spin representations.  In the limiting case \eqref{single-spin} considered here, the volume dynamics is entirely dictated by the number of GFT quanta via $V(\phi) = V_{j_o} N_{j_o}(\phi)$; this is essentially identical to the heuristic interpretation suggested by the LQC `improved dynamics' relating LQC to the underlying LQG spin networks.  Finally, note that the missing connectivity information in the simple GFT condensates considered here does not play any role in LQC either.

Of course, if only one mode $j=j_o$ contributes to the effective dynamics, then the correct classical limit requires a milder condition on the microscopic dynamics to reproduce the classical Friedmann equation with respect to the more general case considered in the previous subsection, namely that $m_{j_o}^2 = 3 \pi G$ (and there are no requirements on the other $m_j$).

Therefore, we set $m_{j_o}^2 = 3 \pi G$ in the following so that the correct classical limit is ensured. Now, since $\rho_j = 0$ for all $j \neq j_o$, only $Q_{j_o}$ is non-zero and
\be
\pi_\phi = \hbar Q_{j_o} .
\ee
Given \eqref{single-spin}, the total volume is simply given by
\be
V = V_{j_o} \rho_{j_o}^2 ,
\ee
and the first modified Friedmann equation simplifies to
\be
\left( \f{V'}{3 V} \right)^2 = \f{4 \pi G}{3} + \f{4 V_{j_o} E_{j_o}}{9 V} - \f{4 V_{j_o}^2 \pi_\phi^2}{9 \hbar^2 V^2} ,
\ee
which can be rewritten, using the relation for the energy density of a massless scalar field $\rho = \pi_\phi^2 / 2 V^2$, as
\be \label{fr-ss1}
\left( \f{V'}{3 V} \right)^2 = \f{4 \pi G}{3} \left( 1 - \f{\rho}{\rho_c} \right) + \f{4 V_{j_o} E_{j_o}}{9 V} ,
\ee
with $\rho_c = 3 \pi G \hbar^2 / 2 V_{j_o}^2 \sim (3 \pi / 2 j_o^3) \rho_{\rm Pl}$.  It is clear that the first term is the classical limit, and that the second term is a quantum gravity correction.  In addition, from scaling arguments (the Friedmann equation must be invariant under $V \to \alpha V$ and $\pi_\phi \to \alpha \pi_\phi$) and dimensional analysis, it follows that $E_{j_o} \sim \sqrt{G} \, \pi_\phi / \hbar$.  Since $V_{j_o} \sim \lp^3 \sim \hbar^{3/2}$, it follows that $V_{j_o} E_{j_o} \sim \sqrt\hbar$ and the third term is also a quantum gravity correction to the classical Friedmann equation.

Similarly, the second modified Friedmann equation is
\be \label{fr-ss2}
\f{V''}{V} = 12 \pi G + \f{2 V_{j_o} E_{j_o}}{V} .
\ee
Since the last term here also vanishes in the limit of $\hbar \to 0$, it is also to be understood as coming from quantum gravity corrections.

By comparing these modified Friedmann equations for the isotropic GFT condensate \eqref{cond-wf}, under the condition \eqref{single-spin}, to the effective Friedmann equations of loop quantum cosmology given in Appendix~\ref{app-lqc}, it is clear that they are identical in the case that $E_{j_o} = 0$, while there are differences of order $\sqrt{\hbar}$ if $E_{j_o}$ is not zero.  Therefore, the condensate states considered here reproduce, within the full GFT theory, a similar type of quantum-corrected cosmological dynamics as the ones found in LQC.  In fact, even the form of $\rho_c$ obtained here is closely analogous to that of the critical energy density of LQC, which according to the heuristic relation between LQC and LQG also goes as $\sim \rho_{\rm Pl} / j_o^3$.

These results therefore provide strong support for the qualitative results of LQC, obtained through the assumption of a particular heuristic relation between full loop quantum gravity and the minisuperspace models of LQC.  Remarkably, very similar effective dynamics arise for the GFT condensate states that mirror the heuristic relation that lies at the heart of LQC.

A few differences must be stressed as well.  First, GFT condensate states provide a significantly more general framework than the specific LQG states used to motivate various constructions in LQC; for example, it is possible and even straightforward to consider states with contributions coming from many different $j$, as is clear from the discussion in Sec.~\ref{ss.fried}.  Another important difference is the presence of a new quantum number $E_{j_o}$ which corresponds to an `energy' of the GFT condensate in the spin $j$ (with respect to the relational time $\phi$), but whose fundamental geometric interpretation remains, for now, unclear.  However, the effect of $E_{j_o}$ on the dynamics is relatively simple: if $E_{j_o} > 0$ then the bounce occurs at an energy density greater than $\rho_c$, on the other hand, if $E_{j_o} < 0$ then the bounce occurs at an energy density less than $\rho_c$.  Importantly, no matter the value of $E_{j_o}$, a bounce similar to that found in LQC will always occur: the resolution of the big-bang and big-crunch singularities in GFT condensate cosmology is robust to changes in $E_{j_o}$ (of course, so long as the regime $ \rho \approx \rho_c$ falls within the regime of approximations used to obtain the above cosmological dynamics from the full microscopic dynamics).

Note also that these modified Friedmann equations (for single-spin condensates with $E_{j_o} = 0$) are also found to arise in a different quantum gravity theory, namely brane-world cosmology scenarios with an extra time dimension \cite{Shtanov:2002mb}.

A final comment is in order regarding the space-time interpretation of the GFT condensate.  If there are only a few GFT quanta, it is doubtful that it is possible to associate a continuum space-time to such a quantum state, because the approximation between GFT condensate states and continuum geometries should be expected to fail. Moreover, there is no reason to expect that the hydrodynamic approximation used here to extract the effective cosmological dynamics would still be viable if only a few fundamental GFT quanta are excited from the Fock vacuum.  (Although note that the hydrodynamic equations of motion would formally still apply even in this case.)  Thus, a large number of GFT quanta are necessary for the approximations used here to be valid.  Note however that this is not necessarily problematic in a bouncing cosmological context: as is well known in LQC, the bounce occurs when the space-time curvature nears the Planck curvature scale, not when the volume of the universe is comparable to the Planck volume (another way to put it is that it is the Planck energy density that sets the scale of  the bounce, not the Planck volume).  The same is true here: the bounce occurs when the space-time curvature is large, and therefore the space-time can (and typically does) bounce at volumes that are much larger than the Planck volume even though the space-time curvature is Planckian.  When this is the case, there will be a large numbers of fundamental GFT quanta even during the bounce, and therefore, modulo other considerations, there will be room for a space-time interpretation of the GFT condensate at all times, including the bounce point.

\section{The Vacuum Limit}
\label{s.vac}

In principle, it is also an interesting problem to study the effective cosmology of GFT condensates, and of quantum gravity more generally, in absence of matter fields. However, there are several difficulties in doing so (some of which have already been noticed in LQC \cite{MartinBenito:2008wx, MartinBenito:2011fv}).

In the present context, the vacuum limit of the FLRW space-time, which should give Minkowksi space, is non-trivial for a number of reasons.  The main problem is that, in the absence of a matter field, there is no longer a relational clock available and therefore the issue of defining and using diffeomorphism invariant observables (often subsumed in the label `problem of time') in quantum gravity must be faced head-on, as one should not expect to have any reliable physics otherwise.  In particular, the geometric interpretation of the condensate wave function $\sigma(g_v)$ is not clear: if $\sigma(g_v, \phi_o)$ is to be interpreted as a three-dimensional spatial slice at the relational `instant of time' $\phi_o$, then it is not immediately obvious how to extract a four-dimensional space-time, or even a reliable phase space interpretation, from $\sigma(g_v)$. This is because there is no control over which spatial slice it refers to, and any apparent physical feature could be attributed instead to some peculiar slicing of spacetime.

Even starting from the Friedmann equation \eqref{gen-cond-friedmann}, it is not obvious how to obtain the vacuum limit.  While the right-hand side of \eqref{gen-cond-friedmann} is easily handled by simply setting $Q_j = 0$, the left-hand side is much trickier, since the derivative with respect to $\phi$ becomes ill-defined.  The fact that this is a singular limit is obvious from the classical theory, where the Friedmann equation is simply $V' / 3V = 4 \pi G$: there is no straightforward way to take the vacuum limit (at least, not without using the relation \eqref{dt-dphi} which holds in general relativity but cannot be assumed blindly for a GFT condensate state).

Furthermore, we already know that the Gross-Pitaevskii approximation we use is only the simplest truncation available, which may capture some relevant features of the emergent continuum spacetime, as we have shown, but certainly neglects a lot of important aspects of the theory (e.g., connectivity information). While one can argue that these simple condensates encode enough information to reproduce the dynamics of homogeneous cosmologies, how much is actually captured can only be checked a posteriori, by studying the emergent cosmological dynamics, as we have done. Then, recovering the Friedmann equations in an appropriate semi-classical limit provides strong evidence that in fact neglecting some of the microsopic details like the connectivity information of the spin network nodes was a reasonable approximation.  However, in Minkowski space there is no dynamical equation that can be recovered in order to provide this type of check.

An additional difficulty with the Gross-Pitaevskii condensate wave function for a vacuum GFT model of EPRL type ---in which case there is no derivative operator on the group manifold in the kinetic term of the GFT action--- is that the interactions cannot be negligible.  This is obvious from the equation of motion for the (isotropic) condensate wave function for a vacuum GFT model, which has the general form
\be \label{eom-vacuum}
\sigma_j + w_j \bar\sigma_j^4 = 0.
\ee
Clearly, for this equation to hold, the amplitude of the interaction term must be the same as that of the linear term.  Therefore, the interactions in this case are necessarily large, so we have to work exactly in the regime where the Gross-Pitaevskii approximation cannot be justified from the microscopic theory.  For the vacuum case, it is then anyway necessary to consider more complicated condensate states, perhaps such as those constructed in \cite{Oriti:2015qva}. 

One can still learn some interesting lessons about GFTs and spin foam models from studying these equations. For example, any solution of these equations is necessarily non-perturbative in the GFT coupling constant (if only one interaction is included), and thus it is not reproducible (nor is its putative physics) via the spin foam expansion. This is in fact the case for all the classical solutions to GFT models that have been studied to date (see e.g., \cite{emergentmatter,emergentmatter2}). However, for the reasons we have explained above, it is unclear how these formal insights can be turned into physical ones.

\section{Discussion}
\label{s.disc}

In this paper, we started from a rather general definition of the EPRL spin foam model for the microscopic dynamics of quantum geometry in the GFT framework (completing the spin foam definition). We parametrized some of the ambiguities entering the definition of the model, by working with general kinetic kernels, and with the aim of constraining these ambiguities  by studying the effective cosmological dynamics they give rise to. While we focused on one specific model, we expect many of our results to hold more generally, and there are indeed several indications that this is the case.

Then, we explained how to couple quantum geometry with a free massless scalar field and gave the definition of the corresponding extension of the GFT model.  One of the key steps here was enforcing at the GFT level the basic symmetries characterizing the scalar field path integral in the free massless case and then performing a small derivative expansion keeping only the lowest orders.  This gives a relatively simple model that we are confident captures the correct semi-classical continuum physics.  At the same time, this allows us to parametrize some ambiguities by working with more general coefficients encoding the matter-geometry coupling.  This is our first new result.  Beside the intrinsic interest of coupling matter and geometry in a GFT model, it was pivotal for the study of the effective cosmological dynamics since it allows for a formulation of the evolution of the universe in terms of purely diffeomorphism invariant observables, as the theoretical context demands, by using the scalar field as a relational clock; this is also a traditional strategy in quantum cosmology.  Interestingly, the scalar field degrees of freedom also enter the GFT action in the same way in which a local time variable would do in standard quantum field theory.

The next step was to implement a restriction to purely isotropic degrees of freedom for the quantum geometric condensate states to be used for the extraction of cosmological dynamics.  These already implement a notion of homogeneity, as discussed in Sec.~\ref{s.cond}, and are thus adapted for the simplest cosmological space-times.  We define what it means to restrict the condensate wave function to isotropic configurations by analysing the simplicial geometry of the GFT states. Indeed, pure volume information can be most directly extracted by the use of quantum states whose quanta can be described by equilateral tetrahedra. This is our second, intermediate, result.

Armed with the resulting simple condensate states, their effective cosmological dynamics follow directly from the full quantum dynamics of the model, in the GFT analogue of a Gross-Pitaevskii approximation (amounting to the simplest mean field treatment), and correspond to the hydrodynamics of the GFT model.  Despite the simplicity of the GFT condensate states, the emerging cosmological dynamics are extremely interesting.  The equations of motion are non-linear with respect to the condensate wave function, and they can be interpreted in analogy with a quantum cosmology wave function.  Thus, this equation has the form of an extension of a quantum cosmology dynamics as used, e.g., in loop quantum cosmology, and with analogous geometric variables.  Due to the isotropic restriction on the condensate wave function, the actual form of the equation simplifies greatly and it can be manipulated in a straightforward manner.

By defining appropriate relational volume observables, we obtained two equations governing the dynamics of the volume of the universe as a function of the massless scalar field, and depending on the particular solution of the effective GFT condensate hydrodynamics through the values of $E_j$ and $Q_j$.  These two new quantities correspond to state-dependent conserved charges of the system (in the Gross-Pitaevskii approximation, and when interactions are subdominant).  The two equations take the form of modified Friedmann equations and are our third main result.

Their interpretation is confirmed by their subsequent analysis. First of all we have shown that in an appropriate semi-classical limit they reduce to the standard Friedmann equations of general relativity. This is true for any state chosen (among the solutions of the GFT condensate dynamics) and at all times (i.e., for all values of the scalar field) if and only if precise conditions on the initial kinetic kernels of the microscopic GFT models are satisfied. These conditions also provide a definition of Newton's constant $G$ as a function of the microscopic parameter of the theory, as expected in such a hydrodynamics context.  This is our fourth main result.

Beyond this classical approximation, the theory provides a number of quantum corrections to the classical Friedmann dynamics. All of them can be computed from first principles starting from the microscopic GFT theory. We have shown that they share one common feature: the cosmological singularities predicted by the classical theory are generically avoided at least for Gross-Pitaevskii condensate states and a quantum gravity bounce takes their place.  This is our fifth main result.

The scenario suggested by the GFT condensate ansatz for cosmological dynamics is therefore similar to the one found in loop quantum cosmology, which is therefore corroborated by the results found here.  In fact, further evidence in support of LQC is obtained by considering GFT condensate states chosen to only have support on one spin $j$ --- this is the type of state that LQC suggests is relevant for the cosmological sector of LQG.  For this type of condensate state, the resulting effective cosmological dynamics also simplify greatly and can be analyzed in detail, with the result that the effective dynamics are extremely similar to the LQC effective Friedmann equations.  This is our last main result.

\bigskip

Obviously, several approximations were required in order to obtain the results summarized above from the full theory.  As a consequence, these results can only be trusted so long as the GFT state remains within the domain of validity of these approximations.  Let us point them out, for completeness.

First, aside from the choice of the microscopic GFT model itself, we used a small derivative approximation with respect to the scalar field, which is then assumed to be slowly varying throughout the evolution of the universe. 

Second, even assuming that the relevant quantum states to study cosmological dynamics within the full GFT theory are condensate states, we used a very simple type of such states, i.e., coherent states of the GFT field operator (i.e., the Gross-Pitaevskii condensate states).  This is a drastic approximation; generic condensate states are much richer.  In particular, we have neglected to include any information about the connectivity of the fundamental spin network degrees of freedom. While the connectivity information is not required in order to characterize homogeneous and isotropic space-times, this nevertheless represents an important restriction.  For example, it implies that the only information about the topology of the universe is the one that can be inferred from the effective cosmological dynamics itself and that the topology of the space-time cannot be constrained at level of the quantum states.  Still, as is well known from the study of real Bose condensates, even these drastic simplifications in the choice of the condensate states that are considered can provide a wealth of information about the relevant physics.

The restriction to isotropic degrees of freedom is of course an additional approximation.  While it does restrict the GFT condensate states to only depend on the degrees of freedom of interest, it has the disadvantage of being imposed as a starting point and for this reason it may prevent the understanding of some aspects of cosmological dynamics and in particular it makes it impossible to study how an isotropic configuration may be generated from a more general initial state.

Given these approximations in the definition of the GFT model and the choice of the condensate states, we then extracted the effective cosmological dynamics directly from the full quantum dynamics of the microscopic GFT theory.  However, the equation we used is the first Schwinger-Dyson equations, corresponding to the classical GFT equations of motion, and is the simplest formulation of the hydrodynamics of the theory.  Ignoring the other Schwinger-Dyson equations is a truncation of the full quantum dynamics.

Having obtained the equations of motion for the condensate wave function, we further assumed that the GFT interaction terms (corresponding to the non-linear terms in the condensate hydrodynamics) in this equation are subdominant compared to the GFT kinetic terms.  This is the case for small values of the coupling constants (which are subject to the renormalization group flow) and for state densities that are not too large.  This is the direct GFT counterpart of the dilute gas approximation used for standard Bose liquids in the Gross-Pitaevskii approximation.  In fact, this approximation is required for consistency with the previous approximation that connectivity information is negligible, since stronger interactions would imply stronger correlations between the GFT quanta, in which case it would become necessary to take into account the connectivity between the spin network nodes and to work with more complicated condensate states.

A last type of approximation was not explicitly stated since it does not enter directly any of the calculations and it is not in fact required by the formalism itself.  This is the condition that curvatures associated to individual GFT quanta are not too big and that their individual contribution to the total volume is rather small. This is required for a straightforward interpretation of the corresponding quantum states as associated to a continuum geometry.  At this point, it is not yet clear how strictly this condition should be enforced and at what point it fails.  A more careful inspection of explicit solutions may provide some insight in this direction.

Within the regime of the full theory in which the above approximations hold, our results hold as well and, we believe, provide interesting insights into the cosmological sector of LQG and GFT.

\bigskip

This being said, there remain many open questions to be addressed in order to further corroborate these results and also to develop the path they open.  We shall conclude by outlining some of the open issues and directions for future research.

It is important to repeat the analysis done here for the GFT based on the EPRL spin foam model for another GFT based on a different spin foam model for 4D quantum gravity. We do not anticipate very different results, though, due to the fact that the main difference between the different GFT models will be encoded in the GFT interaction terms and we have assumed their contributions to the equations of motion of the condensate wave function to be subdominant.  Therefore, a different interaction term would only give different types of quantum corrections to the general cosmological dynamics we have found.  Moreover, we expressed the kinetic terms of the GFT action in a rather general form in order to minimize the reliance on the details of the microscopic model.  However, any difference would be worth investigating.  For example, it is not obvious that models enforcing a weaker relation between $SU(2)$ and $SL(2,\mathbb{C})$ data than the EPRL model would give an effective equation as simple as the one we obtained in the isotropic restriction, and this may offer interesting insights on the role of Lorentzian structures at the effective cosmological level.

Also concerning the choice of the fundamental GFT model, a more detailed analysis of the coupling of the scalar field to (quantum) geometry is needed involving a careful analysis of the Feynman amplitudes of the coupled GFT model, and in particular it is necessary to go beyond the approximation of small derivatives.  One would like the GFT model coupled to a free massless scalar field to define a proper simplicial path integral for discrete gravity coupled to a scalar field discretized on the same cellular complex corresponding to the GFT Feynman diagram.  This analysis should also suggest how to go beyond the case of a free massless scalar field.  While a scalar field with a non-trivial potential would not be a good global clock, it would be interesting to understand how generic scalar fields should be treated in GFT models.  Moreover, it would be interesting to see how the new terms encoding its potential at the level of the GFT action modify its role as a time variable for the same field theory.  Going even further, it would also be good to include other types of matter fields like Maxwell and Dirac fields, and also to better understand the vacuum case.

Another important next step is to perform a more detailed analysis of the solutions to the cosmological equations we have obtained.  It is only by having at hand explicit solutions of the dynamics that we will be able to study the detailed evolution of geometric observables (and specifically the spatial volume of the space-time as well as the relational Hubble rate) and obtain a detailed picture of the evolution of the universe as dictated by the fundamental GFT.  In addition, this analysis will also allow us to check the exact regimes of validity of the various approximations we employed.

In particular, a careful analysis of this type will provide a better understanding of the quantum corrections to the classical Friedmann equations as predicted by the GFT, and in particular of those coming from the GFT interaction terms (which could perhaps be expressed in terms of geometric observables).  Clearly, the interaction terms will generate quantum gravity corrections to the classical equations of general relativity which will be interesting to study, and in addition, as already explained, the interaction terms encode the differences between different microscopic definitions of the quantum gravity dynamics (i.e., different spin foam models or GFT actions) can be elucidated.  So this is a further reason to study the effect of the interaction term in the GFT action on the emergent cosmological dynamics.  It would also be interesting to derive an effective field theory that captures the same corrections to classical general relativity for homogeneous and isotropic space-times, since this would be a nice way to characterize the type of effective continuum dynamics of geometry that is predicted by the GFT models.

Another important open question is to understand how to encode spatial curvature in the effective dynamics of GFT condensate states.  More specifically, the Friedmann equations recovered for the condensate states studied in this work are, in the classical limit, those for a spatially flat FLRW space-time.  How might it be possible for GFT condensate states to give the Friedmann equation for FLRW space-times with non-vanishing spatial curvature?  There are two obvious possibilities: either by studying more involved condensate states, or by modifying the GFT dynamics.  Let us comment more on the first possibility: perhaps it is necessary to encode the spatial curvature in the connectivity information of the condensate state.  (This is consistent with ---but certainly not implied by--- the fact that in the condensate states considered here where the connectivity information is discarded, the spatially flat FLRW space-time is automatically recovered.)  If this is the case, then in order to properly answer this question it will be necessary to study GFT condensate states where the connectivity information is included in the analysis, e.g., states of the type constructed in \cite{Oriti:2015qva}.

The isotropic restriction on the condensate wave functions could also be lifted.  This would allow for a larger class of operators on the condensate wave function, and could potentially be used in order to study anisotropic (and homogeneous) space-times.  In that case, removing the isotropic restriction could perhaps also give some insight into the transition between isotropic and anisotropic regimes in full quantum gravity.

Finally, there remain two related important open problems concerning extensions of the study of cosmology using GFT condensate states.  The first is to tackle the regime of the dynamics where the GFT interactions cease to be subdominant and have to be included explicitly in the effective cosmological dynamics.  This is not so much a technical issue (since it would be rather straightforward to include the interaction terms in the equations derived in this paper and solve them in the case when the interactions are large), but a conceptual one since the Gross-Pitaevskii approximation fails to be justified by the microscopic theory when interactions become important.  In fact, in the regime where the interaction term is no longer negligible it will likely be necessary to choose a different class of condensate states and, in particular, it may be necessary to use condensate states which take fully into account the connectivity of the underlying fundamental spin network states.  This can be done, but the degree of complexity of the analysis necessary in order to extract the effective cosmological dynamics will be considerably higher than what was needed here (among other complications, a number of graph theoretic issues will need to be addressed).

At a more physical level, the second longer-term direction of research will be to develop a generalization of GFT condensate cosmology in order to include perturbations over the condensate state representing an homogeneous universe, as this would allow for a treatment of cosmological perturbations.  A first step would be to reproduce in the context of the full GFT formalism the basic set-up of the analysis of cosmological perturbations performed in loop quantum cosmology; in particular, the separate universe picture \cite{WilsonEwing:2012bx, Wilson-Ewing:2015sfx} appears rather straightforward to implement. The real goal, however, is more ambitious: to develop cosmological perturbation theory from first principles in a fundamental theory of quantum gravity, in this case GFTs.

\appendix

\section{Relational Dynamics in Cosmology}
\label{app}

This appendix contains a brief review of the Hamiltonian formalism applied to a spatially flat FLRW space-time, and the resulting Friedmann equations in general relativity and loop quantum cosmology.  As in the main text, we consider the case where the matter content is a minimally coupled massless scalar field $\phi$.

In order to avoid divergent integrals in the Hamiltonian formulation, we assume that the topology of the spatial surfaces is $\mathbb{T}^3$ and that the volume of the 3-torus with respect to the coordinates $\vec{x}$ is 1.

\subsection{General Relativity}
\label{app-gr}

The line element for a flat FLRW space-time is
\be \label{app-line}
ds^2 = - N^2 d\tau^2 + a(\tau)^2 d \vec{x}^2,
\ee
with $N$ being the lapse and $a(\tau)$ the scale factor, and the Hamiltonian constraint ---coming from the Einstein-Hilbert action, given the line element \eqref{app-line}--- is given by
\be \label{c-ham}
{\cal C} = \f{-3 N V H^2}{8 \pi G} + \f{N \pi_\phi^2}{2 V} = 0,
\ee
where $V = a^3$ is the volume, $H$ is the variable canonically conjugate to $V$ with the Poisson bracket $\{H, V\} = 4 \pi G$, and $\pi_\phi$ is the momentum of the scalar field and satisfies $\{\phi, \pi_\phi\} = 1$.

The equations of motion are easily derived from the Hamiltonian; the two that are necessary for our purposes are those for $\phi$ and $V$.  For the massless scalar field,
\be \label{dt-dphi}
\f{d \phi}{d \tau} = \{\phi, {\cal C}\} = \f{N \pi_\phi}{V}, \quad \Rightarrow \quad d \tau = \f{V}{N \pi_\phi} d \phi,
\ee
which is clearly monotonic since $\pi_\phi$ is a constant of motion as $\{\pi_\phi, {\cal C}\} = 0$.  Thus, the scalar field can act as a good relational clock.  The dynamics of $V$ is given by
\be
\f{d V}{d \tau} = \{V, {\cal C}\} = 3 N V H, \quad \Rightarrow \quad \f{1}{3 V} \, \f{d V}{d \tau} = N H,
\ee
showing that $H$ is the Hubble rate in proper time.  Then, using the vanishing of the Hamiltonian constraint \eqref{c-ham} and rewriting the time derivative in terms of the massless scalar field via \eqref{dt-dphi} gives the first Friedmann equation in terms of the relational clock $\phi$,
\be \label{fr-1}
\left( \f{1}{3 V} \, \f{d V}{d \phi} \right)^2 = \f{4 \pi G}{3}.
\ee

It can be checked that the continuity equation, once the energy density and pressure have been expressed as $\rho = P = \pi_\phi^2 / 2 V^2$, is simply $d \pi_\phi / d \phi = 0$ which is trivially satisfied since $\pi_\phi$ is a constant of the motion and therefore the remaining relevant equation of motion is the second Friedmann equation obtained by differentiating \eqref{fr-1} with respect to $\phi$, giving
\be \label{fr-2}
\f{1}{V} \, \f{d^2 V}{d \phi^2} = \f{1}{V^2} \left( \f{d V}{d \phi} \right)^2 = 12 \pi G.
\ee

The equations \eqref{fr-1} and \eqref{fr-2} are the two Friedmann equations for a spatially flat FLRW space-time expressed in terms of the relational clock $\phi$, which can be directly compared to the Gross-Pitaevski equation of the condensate states studied in this paper.

\subsection{Loop Quantum Cosmology}
\label{app-lqc}

Using the non-perturbative quantization techniques of loop quantum gravity, in loop quantum cosmology (LQC) it has been possible to obtain a well-defined quantum theory for FLRW space-times where the big-bang and big-crunch singularities are resolved \cite{Bojowald:2001xe, Ashtekar:2006wn, Ashtekar:2007em}.  An important result in LQC (and in quantum cosmology in general) is that states that are semi-classical (i.e., sharply peaked around a classical solution) in the low curvature regime remain sharply-peaked throughout their evolution, for the reason that observables in the quantum cosmology are global quantities and hence are heavy degrees of freedom \cite{Rovelli:2013zaa}.

For these semi-classical states, to determine the main features of the quantum dynamics it is sufficient to calculate the evolution of the expectation values of only a few observables, for example the total volume $V$ and the momentum of the scalar field $\pi_\phi$.  Interestingly, the dynamics of these observables, for sharply peaked states, are given by a simple modification to the Friedmann equations.  These are called the LQC effective Friedmann equations, and in terms of proper time $t$ (i.e., for $N=1$) they are \cite{Taveras:2008ke}
\be
H^2 = \f{8 \pi G}{3} \rho \left( 1 - \f{\rho}{\rho_c} \right),
\ee
and
\be
\d H = -4 \pi G (\rho + P) \left( 1 - \f{2 \rho}{\rho_c} \right),
\ee
where $H = \d V / 3 V$ is the Hubble rate, a dot denotes a derivative with respect to proper time, and $\rho_c \sim \rho_{\rm Pl}$ is the critical energy density of LQC.  From these equations it is clear that a bounce in the scale factor occurs when the energy density of the matter field equals the critical energy density $\rho_c$, and hence in LQC a contracting universe bounces into an expanding universe, with a quantum gravity bridge linking the two classical branches of the space-time.

Also, from these equations it can easily be checked that the continuity equation is unchanged,
\be
\d\rho + 3 H (\rho + P) = 0.
\ee
Recall that for a massless scalar field, $\rho = P = \pi_\phi^2 / 2 V^2$ and, as in the case of general relativity, the continuity equation reduces to $\d\pi_\phi = 0$.

In order to make contact with the equations of motion for the GFT condensate, it is necessary to express the LQC effective Friedmann equations in terms of the relational clock $\phi$.  (Note that the continuity equation $\d\pi_\phi = 0$ shows that $\phi$ is a good global clock here.)  This can be done via the relation \cite{Taveras:2008ke}
\be
dt = \f{V}{\pi_\phi} d\phi,
\ee
which is unchanged from the classical relation.

Therefore, the LQC effective Friedmann equations in terms of the relational clock $\phi$ are
\be
\left( \f{1}{3 V} \cdot \f{d V}{d \phi} \right)^2 = \f{4 \pi G}{3} \left( 1 - \f{\rho}{\rho_c} \right),
\ee
and
\be
\f{1}{V} \cdot \f{d^2 V}{d \phi^2} = 12 \pi G.
\ee
Interestingly, when expressed in terms of the relational clock $\phi$, only the first LQC effective Friedmann equation is modified from its classical form.  This is simply due to what is effectively a choice of the lapse $N \sim V$ when $\phi$ is used as a relational clock.

\acknowledgments

We would like to thank the participants of the ``Cosmology from Quantum Gravity'' workshop, and in particular Mairi Sakellariadou, Martin Bojowald, Jakub Mielczarek, Steffen Gielen and Andreas Pithis, for helpful discussions.
This work was partially supported by the John Templeton Foundation through the grant PS-GRAV/1401.


\small
\raggedright

\end{document}